\documentclass[11pt,a4paper]{article}
\usepackage{jheppub}
\usepackage{mathrsfs}
\usepackage{amsfonts}
\usepackage{mathtools}
\usepackage{setspace}
\usepackage{cellspace}
\usepackage{amsmath,amssymb,bm}
\usepackage[colorlinks=true,linkcolor=blue]{hyperref}
\usepackage{xcolor}
\usepackage{epsfig}
\usepackage{slashed}
\usepackage{caption}
\usepackage{hhline,multirow,tabularx}  
\usepackage{dcolumn}    
\usepackage{url}        
\usepackage{braket}     
\renewcommand{\bra}[1]{\left<#1\left|}
\renewcommand{\ket}[1]{\right|#1\right>}
\preprint{CNF-UMD-2022}

\title{Generalized parton distributions through universal moment parameterization: zero skewness case}

\author[a]{Yuxun~Guo}
\author[a,b]{, Xiangdong~Ji}
\author[b]{and Kyle~Shiells}
\affiliation[a]{Maryland Center for Fundamental Physics, Department of Physics, University of Maryland,\\ 4296 Stadium Dr., College Park, MD 20742, USA}
\affiliation[b]{Center for Nuclear Femtography, SURA,\\ 1201 New York Ave. NW, Washington, DC 20005, USA}
\emailAdd{yuxunguo@umd.edu}
\emailAdd{xji@umd.edu}
\emailAdd{kshiells@sura.org}
\abstract{We present a global analysis program for the generalized parton distributions (GPDs) based on conformal moment expansion. We apply the strategy of universal moment parameterization to fit both the collinear parton distribution functions (PDFs) from phenomenology and generalized form factors from lattice calculations, and show that the parameterization is flexible enough to accommodate these constraints. In addition, we can also fit direct lattice calculations of GPDs from large-momentum effective theory. In this work we focus on the analysis of $t$-dependent PDFs which correspond to GPDs in the $\xi \to 0$ limit. The strategy also applies to the $\xi \not =0$ region with extra parameters, and therefore can be fitted to experimental observables in the future. With a demonstrative example of fitted GPDs, we exhibit the quark transverse angular momentum densities of the proton as well as the impact parameter space distributions of quarks in both unpolarized and transversely polarized protons.}
\keywords{Generalized parton distributions; Generalized form factors; GUMP;}
\date{\today}
\begin{document}
\maketitle

\section{Introduction}

Generalized parton distributions (GPDs)~\cite{Muller:1994ses, Ji:1996ek, Ji:1998pc} which describe the parton distributions in the nucleon with transverse displacement contain important information about the nucleon such as the mass, angular momentum and mechanical properties~\cite{Ji:1994av, Ji:1996ek, Polyakov:2002yz} as well as the 3-dimensional (3D) structure of the nucleon~\cite{Burkardt:2000za,Burkardt:2002hr,Ji:2003ak, Belitsky:2003nz}. There have been experimental measurements of deeply virtual Compton scattering (DVCS) \cite{Ji:1996nm} cross-sections aiming to explore GPDs at HERA (H1 \cite{H1:2001nez,H1:2005gdw,H1:2007vrx,H1:2009wnw}, ZEUS~\cite{ZEUS:2003pwh,ZEUS:2008hcd}, HERMES~\cite{HERMES:2012gbh,HERMES:2012idp}) and Jefferson Lab (CLAS~\cite{CLAS:2007clm,CLAS:2008ahu,Niccolai:2012sq,CLAS:2015bqi,CLAS:2015uuo,CLAS:2018bgk,CLAS:2018ddh,CLAS:2021gwi} and Hall A~\cite{JeffersonLabHallA:2006prd,JeffersonLabHallA:2007jdm,JeffersonLabHallA:2012zwt,Georges:2017xjy,JeffersonLabHallA:2022pnx}) spanning decades. In addition, lattice quantum chromodynamics (QCD) provides us another opportunity to study the parton physics in the nucleon from first-principles calculations. The moments of GPDs are calculated on the lattice for the quark~\cite{LHPC:2007blg,Hagler:2009ni} and the gluon~\cite{Shanahan:2018nnv}. In recent years, the large momentum effective theory (LaMET)~\cite{Ji:2013dva,Ji:2014gla,Ji:2020ect} was proposed, which expands lattice technology further to allow the direct calculations of the $x$-dependence of the parton distributions including GPDs \cite{Alexandrou:2020zbe,Constantinou:2020hdm,Lin:2020rxa,Lin:2021brq}.

Despite the efforts put into investigating GPDs, limited knowledge of them has been obtained at this point due to their higher-dimension nature. More specifically, GPDs, collectively denoted as $F$, are functions of $(x,\xi,t)$ with $x$ the parton momentum fraction, $\xi$ the skewness parameter describing the momentum transfer in the longitudinal direction and $t$ the total momentum transfer squared. Typically, we have constraints from: parton distribution functions (PDFs) in the forward limit $\xi=t=0$, generalized form factors (GFFs) with $t$-dependence and Compton form factors (CFFs) with $\xi$- and $t$-dependence. None of these place 3D constraints, causing a huge gap in the determination of GPDs. 

Due to the lack of knowledge, we are faced with the so-called inverse problem when reconstructing GPDs, which states that GPDs can not be uniquely determined with, for instance, the combination of PDFs and the CFFs \cite{Bertone:2021yyz}. 
Therefore, we are in need of a global analysis program that can combine different constraints on GPDs to solve the inverse problem. One possible way is to parameterize the GPDs with finite (and extendable) numbers of free parameters, such that one can match the number of free parameters to the number of constraints on the GPDs till they are completely determined with controlled uncertainties. This is 
in the same spirit of determining proton collinear PDFs from various experimental data~\cite{Hou:2019efy,Xie:2021equ}.

With the above motivation, lots of GPD parameterization methods are proposed in the literature~\cite{Radyushkin:1998bz,Radyushkin:1998es,Polyakov:2002wz,Guidal:2004nd,Goloskokov:2005sd,Mueller:2005ed,Kumericki:2007sa,Kumericki:2009uq,Goldstein:2010gu,Gonzalez-Hernandez:2012xap,Kriesten:2021sqc}. They generally fall into three categories: the double distribution representations, the moment space representations and other dynamic GPD models. The Radyushkin double distribution ansatz (DDA) is one of the earliest GPD parameterization method \cite{Radyushkin:1998bz,Radyushkin:1998es}. Under the double distribution framework, the Vanderhaeghen-Guichon-Guidal (VGG) model~\cite{Vanderhaeghen:1999xj,Goeke:2001tz,Guidal:2004nd} and the Goloskokov-Kroll (GK) model ~\cite{Goloskokov:2005sd,Goloskokov:2007nt} were developed for phenomenological applications. In general, the DDA expresses the GPD in terms of an integral expression of the double distribution. Then by constructing proper double distributions, the GPD will be reduced to the corresponding PDF in the forward limit and satisfies the polynomiality conditions. However, since the original double distributions have only one free parameter to control the overall $\xi$ dependence, such GPDs parameterization methods lack the flexibility when fitting to measurements with different $\xi$ \cite{Freund:2002qf}.

The limitation of DDA leads to explorations of other possibilities. Another approach to constructing GPDs is to consider the polynomial expansion of GPDs. Two major GPD parameterization methods based on the polynomial expansion are the dual parameterization \cite{Shuvaev:1999fm,Noritzsch:2000pr,Polyakov:2002wz} and the Kumeri\v{c}ki-M\"uller (KM) model \cite{Mueller:2005ed,Kumericki:2009uq} commonly used in DVCS analysis, which are shown to be mathematically equivalent in ref. \cite{Muller:2014wxa}. The basic idea of the polynomial expansion or the moment expansion is to expand the GPD in terms of a complete set of polynomials where the expansion coefficients are the corresponding moments. Then the GPD can be expressed as the sum of these polynomials without loss of generality. The polynomial expansion requires the polynomials to properly imitate the behavior of the GPD and extra care will be needed for practical uses.

In addition, there are various dynamic models of GPDs \cite{Goldstein:2010gu,Gonzalez-Hernandez:2012xap,Kriesten:2021sqc,Kumar:2017dbf,Liu:2022fvl}. These models generally approximate the hadron states as expansions in the Fock space of quarks and gluons, usually with the help of light front wave functions. Since the parton distributions are the matrix elements of certain light-cone operators between hadrons, they are calculable when the hadron states are replaced with the  bound states of quarks and gluons with some effective interactions. In this sense, the parameterization of GPDs is turned into the task of determining the free parameters in the wave functions and the effective interactions.

However, a general parameterization strategy that combines various physical constraints on GPDs with the lattice calculations still seems to be lacking in the literature. Therefore, in this work we present a program to parameterize GPDs through universal moment parameterization (GUMP), which can be fitted to lattice calculations, PDFs and experimental measurements of CFFs altogether and help solve the inverse problem. We will continue exploring the moment expansion, especially the conformal moment expansion, for reasons to be explained. 
Our work here is closely related to the KM model, but 
the goal of our program is quite different. The KM model aims to just fit to the experiment data, and thus only the sea distributions are parameterized by the general moment expansion, whereas the valence distributions are given by some functional forms on the crossover line only. More efforts are put into the sea rather than the valence distributions in the KM model because experiments such as DVCS can only measure the CFFs, which are mostly from the sea distributions that dominate at small $x$ or $\xi$. As for this program aiming to combine all the possible constraints on GPDs including both experiments and lattice calculations, both sea and quark distributions are important. Therefore, a general extension of the previous parameterization methods to all GPDs is necessary in order to combine the different constraints.

The organization of the paper is as follows. In section \ref{sec:confmom}, we discuss the physical properties of GPDs and introduce the well-established conformal moment expansion of GPDs. In section \ref{sec:modelmom}, we make our assumptions based on phenomenological consideration. In section \ref{sec:tpdf} we apply them to build the $t$-dependent PDFs and study the nucleon 3D structures with the fitting results. In the end, we conclude in section \ref{sec:conc}.

\section{Review on conformal moment expansion of GPDs}

\label{sec:confmom}

In this section, we will introduce the general technique and review salient features of conformal moment expansion for GPDs~\cite{Mueller:2005ed}, which will be our basis for the GUMP program. We start with the physical constraints on GPDs which are important guides to finding proper parameterization strategies.

\subsection{Physical constraints on GPDs}

GPDs are simply functions $F$ with 3 arguments $(x,\xi,t)$, which can be defined as the off-forward matrix element of light-cone operators as,
\begin{equation}
    F(x,\xi,t)\equiv \int \frac{\text{d} \lambda}{2\pi} e^{i\lambda x} \bra{P'} O(\lambda n) \ket{P}\ ,
\end{equation}
with $n$ the light-cone vector $(n^2=0)$, $O(\lambda n)$ the light-cone operators of the quark/gluon and $P$/$P'$ the initial/final nucleon momenta. We define the average nucleon momentum $\bar P\equiv (P+P')/2$, the momentum transfer $\Delta \equiv P'-P$ and the momentum transfer square $t\equiv \Delta^2$. Besides, the skewness parameter $\xi$ is defined as $\xi=-n \cdot \Delta /( 2 n \cdot \bar P)$. In the forward limit, GPDs are reduced to the corresponding PDF $f(x)$, so we have 
\begin{equation}
    \lim_{\xi,t \to 0}F(x,\xi,t)= f(x)\ ,
\end{equation}
which is the most important and stringent constraint on GPDs.

\begin{figure}[t]
\centering
\includegraphics[width=0.7\textwidth]{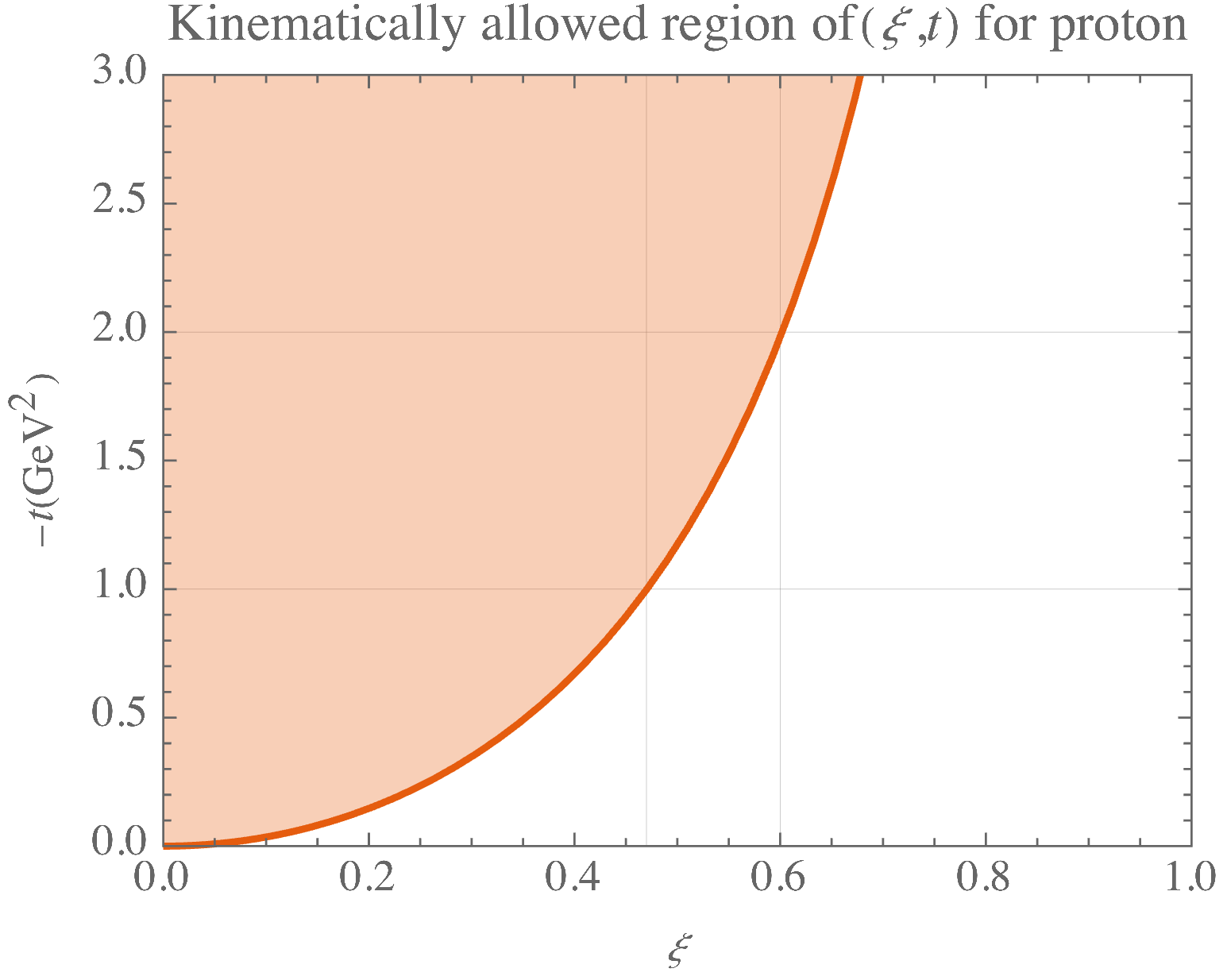}
\caption{\label{fig:txiplot} A plot showing the kinematically allowed region on the $(\xi,t)$ plane for proton scattering with proton mass $M=0.938\text{ GeV}$. Typically, larger momentum transfer $|t|$ is needed to reach the region with larger skewness $\xi$.}
\end{figure}

The two arguments $x$ and $\xi$ of GPDs have support $x,\xi\in [-1,1]$, whereas the momentum transfer squared $t$ could take any negative value $t<0$. Since GPDs are typically defined symmetric in $\xi$, we will consider only positive $\xi>0$ without loss of generality. GPDs do not exist everywhere in the above 3D subspace. For instance, since the skewness parameter $\xi$ corresponds to the longitudinal projection of the momentum transfer, there exists a minimum $|t|_{\rm{min}}$ for each $\xi$ as
\begin{equation}
    |t|\ge|t|_{\rm{min}}= \frac{4 \xi^2 M^2}{1-\xi^2} \ ,
\end{equation}
with $M$ the nucleon mass such that it takes infinite momentum transfer $|t|\to \infty$ to reach $|\xi| \to 1$, as shown in figure \ref{fig:txiplot}. Accordingly, we restrain ourselves to the kinematically allowed region.

Furthermore, there exists non-analyticity on the $(x,\xi)$ plane,  specifically on the crossover line $|x|=\xi$. Since GPDs are essentially parton distributions with non-zero momentum transfer $\Delta$, their physical interpretation varies accordingly~\cite{Ji:1998pc}: with small momentum transfer in the near forward region $\xi<|x|$, GPDs can still be interpreted as the amplitudes of emitting and reabsorbing a parton in the nucleon and behave like the PDFs, whereas with large momentum transfer in the region $\xi>|x|$, they are interpreted as the amplitudes of emitting/absorbing a pair of partons and behave like the distribution amplitudes (DAs). The two regions will be referred to as the PDF region ($\xi<|x|$) and the DA region ($\xi>|x|$) respectively, and they are joined together at the crossover line $\xi=|x|$ where one of the partons has zero momentum fraction. The collinear factorization requires the GPDs to be continuous at the crossover lines, but their derivatives might not be. Consequently, these crossover lines can be extremely subtle for the parameterization of GPDs on the $(x,\xi)$ plane.

\begin{figure}[t]
\centering
\includegraphics[width=\textwidth]{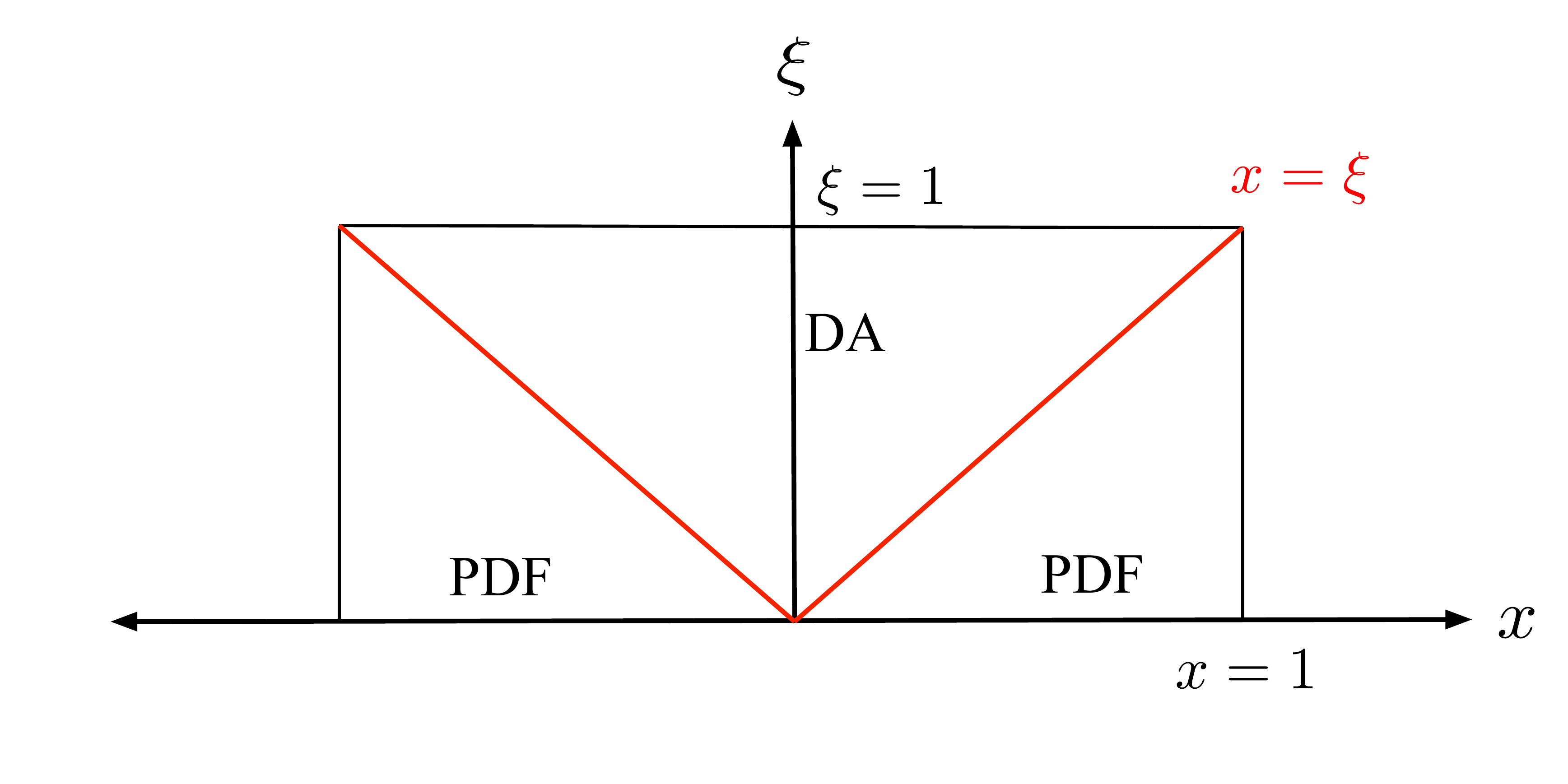}
\caption{\label{fig:extractedCFFsplot} Graphical representation of the kinematical regions of GPDs on the $(x,\xi)$ plane. The GPDs need to be continuous at the crossover line $|x|=\xi$ in order for factorization to work, while their derivatives might not exist there. }
\end{figure}

In addition, GPDs are also constrained by the polynomiality condition, which states that the Mellin moments of GPDs 
\begin{equation}
\label{eq:MelMoment}
    F_{n}(\xi,t)\equiv \int_{-1}^{1} \text{d}x x^{n-1} F(x,\xi,t)\ ,
\end{equation}
must be finite order polynomials of $\xi$ such that \cite{Ji:1998pc},
\begin{equation}
    F_{n}(\xi,t)=\sum_{k=0,{\rm{even}}}^{n} \xi^{k} F_{n,k}(t)\ .
\end{equation}
The polynomiality condition is the result of Lorentz invariance, and it holds even though the integral in eq. (\ref{eq:MelMoment}) involves both the PDF region ($|x|>\xi$) and the DA region ($|x|<\xi$) in spite of the non-analyticity at the crossover lines.

While other physical constraints exist for GPDs, the above constraints will be the main focus of this work.


\subsection{Conformal moment expansion and scale evolution of GPDs}

With the physical constraints discussed above, we now introduce the conformal moment expansion and show how the constraints can be imposed with such expansion. Consider if we expand the GPD in terms of a complete set of polynomials, denoted as $C_{i}(x)$, such that \begin{equation}
\label{eq:polyexpand}
    F(x,\xi,t)=\sum_{i}\rho_C(x)  C_{i}(x) F_{i}^{C}(\xi,t) \ ,
\end{equation}
with $F_{i}^{C}(\xi,t)$  the corresponding coefficient for fixed $\xi$ and $t$. Since $C_{i}(x)$ form a complete set of basis, we have the completeness condition
\begin{equation}
    \sum_{i} C_{i}(x) C_{i}(y)=\rho^{-1}_C(x)\delta(x-y)\ ,
\end{equation}
and the orthonormal condition
\begin{equation}
    \int_{-1}^{1}\text{d} x \rho_C(x) C_{i}(x) C_{j}(x) =\delta_{ij}\ ,
\end{equation}
with $\rho_C(x)$ the weight function of the polynomials $C_{i}(x)$. Then the $F_{i}^{C}(\xi,t)$ is simply the moment of $F(x,\xi,t)$:
\begin{equation}
    F_{i}^{C}(\xi,t) = \int_{-1}^{1} \text{d}x  C_{i}(x) F(x,\xi,t)\ .
\end{equation}
The above is generally true for any complete set of polynomials $C_{i}(x)$ and there are infinite possible choices. Therefore, one naturally would ask: which choice is the most suitable one for GPDs? 
An answer is motivated by the renormalization group equations of GPDs. But before getting into that, we start with a simpler case --- the scale evolution of PDFs, or the Dokshitzer–Gribov–Lipatov–Altarelli–Parisi (DGLAP) equations \cite{Altarelli:1977zs,Dokshitzer:1977sg,Gribov:1972ri}. 

The leading-order DGLAP equations for non-singlet quark distributions reads,
\begin{equation}
\label{eq:DGLAP}
    \frac{\text{d}}{\text{d}\ln Q^2}q(x,Q)=\frac{\alpha_s(Q)}{2\pi}\int_x^1\text{d}y\frac{q(y,Q)}{y}\cdot P\left(\frac{x}{y}\right)~+~\mathcal O(\alpha_s(Q)^2)\ ,
\end{equation}
with $P(z)$ the splitting function. Generally, one needs to solve the integro-differential equation in order to perform the scale evolution of PDFs. However, if one performs the Mellin transform of the DGLAP equation according to
\begin{equation}
\label{eq:Meltrans}
    f_n=\int_{-1}^{1} \text{d}x~x^{n-1} f(x) \ ,
\end{equation}
the DGLAP equation will read,
\begin{equation}
\label{eq:DGLAPMellin}
    \frac{\text{d}}{\text{d}\ln Q^2}q_n(Q)=\frac{\alpha_s(Q)}{2\pi}q_n(Q)\cdot P_n~+~\mathcal O(\alpha_s(Q)^2)\ ,
\end{equation}
in Mellin space, where $q_n(Q)$ and $P_n$ are the Mellin transformations of $q(x,Q)$ and $P(x)$ according to eq. (\ref{eq:Meltrans}) and the DGLAP equation becomes multiplicative in the Mellin moment space which holds true to all order in perturbation theory. Therefore, if one parameterizes the PDFs in the Mellin moment space, the scale evolution equations can be easily solved.

The above moment representation also applies to the construction of GPDs. Consider the scale evolution equations of non-singlet quark GPDs in the form of \cite{Muller:1994ses,Ji:1996nm,Radyushkin:1996nd,Muller:2014wxa,Belitsky:1997pc,Geyer:1982fk,Balitsky:1983sw,Braunschweig:1987dr,Balitsky:1987bk},
\begin{equation}
\label{eq:gpdevolve}
    \frac{\text{d}}{\text{d} \ln Q^{2}} F\left(x, \xi, t, Q^{2}\right)=\frac{\alpha_{s}(Q)}{2 \pi} \int_{-1}^{1} \frac{\text{d} x^{\prime}}{|\xi|}\left[V\left(\frac{x}{\xi}, \frac{x^{\prime}}{\xi}\right)\right]_{+} F\left(x^{\prime}, \xi, t, Q^{2}\right)
\end{equation}
with $\left[V\left(\frac{x}{\xi}, \frac{x^{\prime}}{\xi}\right)\right]_{+}$ the evolution kernel. It can be shown that in the forward limit, the evolution kernel is reduced to the splitting function of DGLAP equation:
\begin{equation}
    \lim_{\xi\to 0 } \frac{1}{|\xi|}\left[V\left(\frac{x}{\xi}, \frac{1}{\xi}\right)\right]_{+} = P(x)\ ,
\end{equation}
while for $|x|,|x'|<1$ it is related to the Efremov-Radyushkin-Brodsky-Lepage (ERBL) kernel \cite{Lepage:1979zb,Lepage:1980fj,Efremov:1978rn,Efremov:1979qk} for the DA scale evolution:
\begin{equation}
    V(2x-1,2y-1)_{0<x,y<1}=V_{\rm{ERBL}}(x,y)\ .
\end{equation}
For this reason, the PDF and DA regions of GPDs have also been called DGLAP and ERBL regions respectively. 

Analogous to the analysis of PDFs and their Mellin moments, the GPD moments that diagonalize the right-hand side (RHS) of eq. (\ref{eq:gpdevolve}) will renormalize multiplicatively. To the leading order, such a solution is found to be the Gegenbauer polynomials  $C_{j}^{\left(\lambda\right)}(x)$, which satisfy \cite{Belitsky:1997pc}
\begin{equation}
    \int_{-1}^{1} \frac{\text{d} x^{\prime}}{|\xi|}\left[V^{(0)}\left(\frac{x}{\xi}, \frac{x^{\prime}}{\xi}\right)\right]_{+} C_{j}^{\frac{3}{2}}\left(\frac{x}{\xi}\right)= \gamma_j C_{j}^{\frac{3}{2}}\left(\frac{x'}{\xi}\right) \ ,
\end{equation}
with $V^{(0)}$ the leading-order evolution kernel and $\gamma_j$ the corresponding anomalous dimension (eigenvalue) of each Gegenbauer polynomial. For the quark we have $\lambda= 3/2$, and for the gluon $\lambda=5/2$ should be used. Accordingly, the conformal moments defined as
\begin{equation}
\label{eq:conformalmoment}
  \mathcal F_{j}(\xi,t)\equiv  \xi^{j} \frac{\Gamma{\left(\frac{3}{2}\right)}\Gamma(j+1)}{2^j \Gamma\left(\frac{3}{2}+j\right)}\int_{-1}^{1}\text{d}x  C_{j}^{\frac{3}{2}}\left(\frac{x}{\xi}\right) F(x,\xi,t)\ ,
\end{equation}
are multiplicatively renormalizable, which will not hold if the higher order evolution effects are taken into account and more care will be needed then \cite{Kumericki:2007sa}. Here an extra normalization constant is multiplied to ensure that in the forward limit, the conformal moments are reduced to the Mellin moments,
\begin{equation}
  \lim_{\xi\to 0}\mathcal F_{j}(\xi,t) =\int_{-1}^{1}\text{d}x ~x^j F(x,\xi,t)\ ,
\end{equation}
which shall be expected as the evolution equation itself will be reduced to the DGLAP equation. Also note that there exists a mismatch of conventions in the literature that the $j$th conformal moment $F_{j}(\xi,t)$ actually corresponds to the $n$th Mellin moment with $n=j+1$ instead of $j$, according to eq. (\ref{eq:Meltrans}). In this work, the two will be distinguished by having $j$ for the conformal moments and $n$ or $s$ for the Mellin moments respectively, and they are related by $n=s=j+1$.

Inspired by the leading-order evolution and with eq. (\ref{eq:polyexpand}), in general GPDs can be written in terms of a formal summation as \cite{Muller:2014wxa},
\begin{equation}
\label{eq:conformalsum}
  F(x,\xi,t) = \sum_{j=0}^{\infty} \xi^{-j-1} \frac{2^j \Gamma\left(\frac{5}{2}+j\right)}{\Gamma{\left(\frac{3}{2}\right)}\Gamma(j+3)} \left[1-\left(\frac{x}{\xi}\right)^2\right] C_{j}^{\frac{3}{2}}\left(\frac{x}{\xi}\right) \mathcal {F}_{j}(\xi,t) \quad \text{for }|x|<\xi\ ,
\end{equation}
where we have the weight function of Gegenbauer polynomials $\rho_C^{\frac{3}{2}}(x)=1-x^2$ and some extra normalization constants to match normalization of the conformal moments $\mathcal {F}_{j}(\xi,t)$ in eq. (\ref{eq:conformalmoment}). The whole prefactors of the conformal moments can be collected to define
\begin{equation}
  (-1)^j p_j(x,\xi)\equiv  \xi^{-j-1} \frac{2^j \Gamma\left(\frac{5}{2}+j\right)}{\Gamma{\left(\frac{3}{2}\right)}\Gamma(j+3)} \left[1-\left(\frac{x}{\xi}\right)^2\right] C_{j}^{\frac{3}{2}}\left(\frac{x}{\xi}\right)\quad \text{for }|x|<\xi\ ,
\end{equation}
where an extra $(-1)^j$ factor is extracted for future convenience. Analogous to the partial wave expansion, the $p_j(x,\xi)$s are called the conformal wave functions and $\mathcal {F}_{j}(\xi,t)$ the conformal partial wave amplitude respectively~\cite{Mueller:2005ed}.

Also note that with the above conformal partial wave expansion, the non-analyticity at $|x|=\xi$ can be taken care of --- GPDs might not have well-defined derivatives there, but the moments still exist such that the polynomiality condition can be imposed.

\subsection{Resummation of conformal partial wave expansion}

In moment expansion, the reconstruction of PDFs or GPDs requires summing over the contributions of all moments. Such summation does not always converge, and analytic methods shall be developed to perform the formal summation. 

Consider the Mellin moment representation of the PDF $f(x)$ for example. In order to express the PDF in terms of its Mellin moments, one can write $f(x)$ in terms of a formal
summation as
\begin{equation}
\label{eq:Mellinseries}
    f(x)=\sum_{n=0}^{\infty} \frac{(-1)^n}{n !} \delta^{(n)}(x) f_{n+1} \ ,
\end{equation}
which is obviously not summable, since the $\delta(x)$ function and its $n$th derivatives $\delta^{(n)}(x)$ are singular. Even though the RHS of eq. (\ref{eq:Mellinseries}) can be shown to be equivalent to the PDF $f(x)$ formally, the expression is of no practical use.
Therefore, the concept of formal summation has to be introduced, which implies that we consider the RHS of eq. (\ref{eq:Mellinseries}) as an abstract mathematical entity regardless of the convergence of the series.

To actually reconstruct the PDF $f(x)$, the inverse Mellin transformation can be employed:
\begin{equation}
\label{eq:invMellin}
    f(x)=\frac{1}{2\pi i} \int_{c-i\infty}^{c+i\infty} x^{-s} f_s  \text{d} s\ ,
\end{equation}
where the Mellin moments $f_s$ are analytically continued in the sense that $s$ can take complex values rather than just positive integers $n$. This method has been adopted for the numerical analysis of PDFs in the literatures \cite{Graudenz:1995sk,Stratmann:2001pb}. Note that the two equations (\ref{eq:Mellinseries}) and (\ref{eq:invMellin}) define the same PDF $f(x)$, though they look totally different.

To resum eq. (\ref{eq:conformalsum}) for GPDs, one obstacle is that the Gegenbauer polynomials are only complete on the interval $[-1,1]$, while the argument of the Gegenbauer polynomials $x/\xi$  in the above expansion can take any real value. For this reason, we restrict the expression eq. (\ref{eq:conformalsum}) to the region $|x|<\xi$, while extension to the PDF region $|x|>\xi$ has to be defined through analytic continuation. This can be done with the Schl\"{a}fli integral, as discussed in ref. \cite{Mueller:2005ed,Muller:2014wxa},
\begin{equation}
  p_{j}(x, \xi)=-\frac{\Gamma(5 / 2+j)}{\Gamma(1 / 2) \Gamma(2+j)} \frac{1}{2 i \pi} \oint_{-1}^{1} d u \frac{\left(u^{2}-1\right)^{j+1}}{(x+u \xi)^{j+1}}\ ,
\end{equation}
where the integration contour encircles the line segment $u\in[-1,1]$. The analytic continued wave functions can then be derived with this integral representation and one gets \cite{Muller:2014wxa},
\begin{equation}
p_{j}(|x| \leq \xi, \xi)=\frac{2^{j+1} \Gamma(5 / 2+j) \xi^{-j-1}}{\Gamma(1 / 2) \Gamma(1+j)}(1+x / \xi)_{2} F_{1}\left(-1-j, j+2,2 \mid \frac{\xi+x}{2 \xi}\right)\ ,
\end{equation}
and 
\begin{equation}
p_{j}(x>\xi, \xi)=\frac{\sin (\pi[j+1])}{\pi} x^{-j-1}{}_{2} F_{1}\left(\begin{array}{c|}
(j+1) / 2,(j+2) / 2 \\
5 / 2+j
\end{array} ~\frac{\xi^{2}}{x^2}\right)\ .
\end{equation}
These wave functions are equivalent to the Gegenbauer polynomials definition of conformal wave functions when $j$ is integer, and they can be used for more general propose, for instance, the resummation of moments. Note that the above conformal wave functions are continuous at $|x|=\xi$ but their derivatives are not, which is consistent with the behavior of GPDs at the crossover lines. 

Two techniques, the Mellin-Barnes integral \cite{Kumericki:2007sa,Kumericki:2009uq} and the Shuvaev-Noritzsch transform \cite{Shuvaev:1999fm,Noritzsch:2000pr} have been used in the literature to reconstruct the GPDs with moments. Both methods are shown equivalent mathematically \cite{Muller:2014wxa}, and we will focus on the former one here. The Mellin-Barnes integral is based on the Sommerfeld-Watson transformation,
\begin{equation}
\sum_{n  = -\infty}^\infty (-1)^n f_n= \frac 1 {2 i} \oint_C \frac {f(z)} {\sin \pi z} \text{d} z\ ,
\end{equation}
assuming $f(z)$ is analytic. The proof is straightforward, as the $\sin(\pi z)$ in the denominator of RHS leads to poles at all integers $z$ with the residue of $\pi (-1)^{n}  f(z)$ and the equation is simply the residual theorem applied to the RHS. Then, the formal summation in eq. (\ref{eq:conformalsum}) can be written in terms of the following Mellin-Barnes integral in the same manner \cite{Kumericki:2007sa,Kumericki:2009uq}
\begin{equation}
\label{eq:MBintegral}
F(x, \xi, t)=\frac{1}{2 i} \int_{c-i \infty}^{c+i \infty} d j \frac{p_{j}(x, \xi)}{\sin (\pi[j+1])} \mathcal{F}_{j}(\xi, t)\ ,
\end{equation}
where the constant $c$ should be carefully chosen to have all the non-negative poles while the poles of conformal wave functions $p_{j}(x, \xi)$ and conformal moments $\mathcal{F}_{j}(\xi, t)$ should be outside the contour, see more discussions in ref. \cite{Muller:2014wxa}. 

\begin{figure}[t]
\centering
\includegraphics[width=\textwidth]{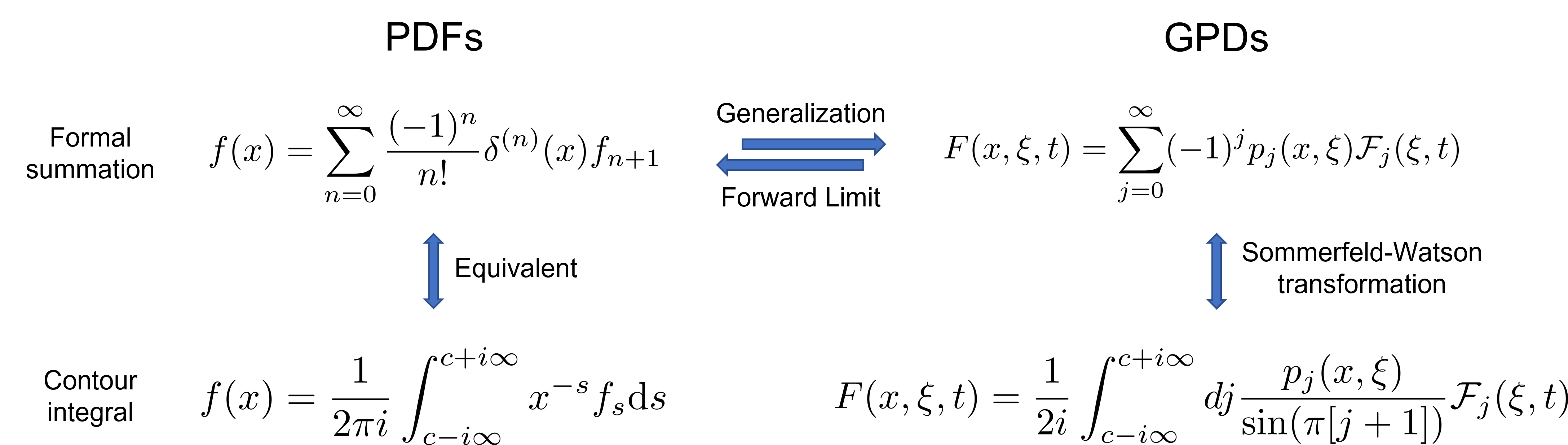}
\caption{\label{fig:analyticmethodcomp}  A comparison of the inverse Mellin transformation expression of the PDFs and the Mellin-Barnes integral expression of GPDs together with the corresponding formal summation expressions.}
\end{figure}

\begin{figure}[t]
\centering
\includegraphics[width=0.6\textwidth]{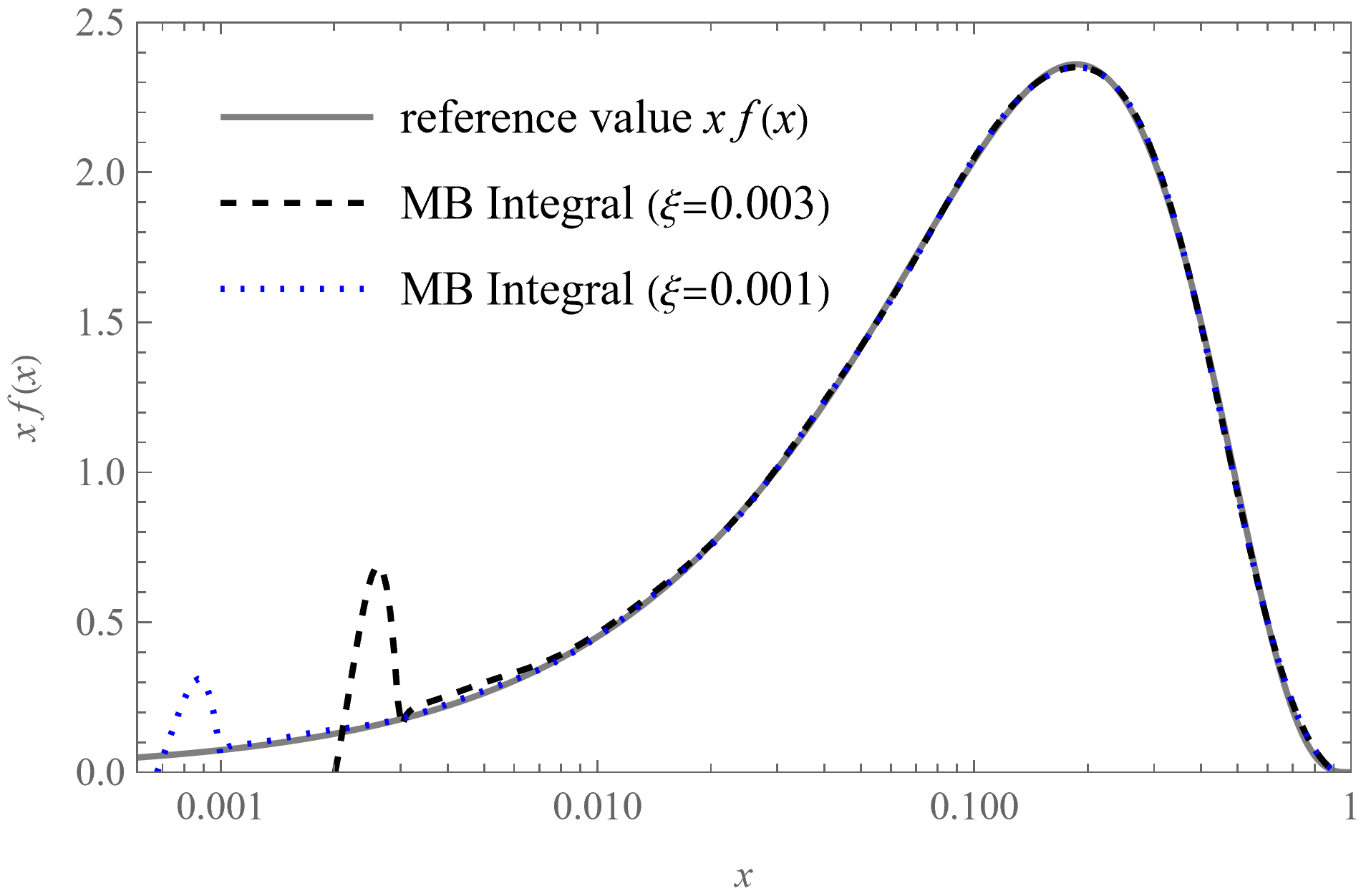}
\caption{\label{fig:mbintegral} A plot showing that the Mellin-Barnes (MB) integral can indeed reproduce the reference PDF $f(x)=18.6 x^{-0.2}(1-x)^{3.5}$ when $x\gg\xi$, where $\xi$ serves as the regulator. On the other hand, in the $x\lesssim \xi$ region, the PDFs calculated this way are strong affected by the none-zero skewness $\xi$ and deviate from the reference value.}
\end{figure}

The comparison of the different methods used for PDFs and GPDs are shown in figure \ref{fig:analyticmethodcomp}. 
In principle, the Mellin-Barnes integral should reproduce the PDFs in the forward limit $\xi=0$. Indeed, since the conformal wave functions $p_j(x,\xi)$s have the limit behavior
\begin{equation}
\lim_{\xi\to 0} p_j(x,\xi) = \frac{1}{j!} \delta^{(j)}(x)\ ,
\end{equation}
the formal summation in eq. (\ref{eq:conformalsum}) will be reduced to eq. (\ref{eq:Mellinseries}). However, due to the singular behavior of $\delta$ functions, the Mellin-Barnes integral will not converge in the $\xi\to 0$ limit. Instead, one can use a finite but small $\xi$ as the regulator for the $\delta$ function, and the Mellin-Barnes integral will reproduce the PDF when $x\gg\xi$, as shown in figure \ref{fig:mbintegral}.

Following the above framework, for any (well-behaved) given set of moments $\mathcal{F}_{j}(\xi, t)$, the corresponding GPDs can be calculated and then used for all kinds of phenomenological applications. The GUMP program for global analysis follows this approach by parametrizing the general forms of the moments or {\it universal moment parametrization}. More details concerning the parameterization of moments will be discussed in the next section.

Note that while the GPDs calculated above are ready for all kinds of applications, there exist some shortcuts, such as the evaluation of CFFs and scale evolution, as these calculations are easier to perform in moment space. The detailed discussion of this topic is in ref. \cite{Kumericki:2007sa}.

\section{Universal parameterization of GPD moments}
\label{sec:modelmom}
In the last section, we show that the GPDs $F(x,\xi,t)$ can be equivalently expressed in terms of their moments $\mathcal F_j(\xi,t)$ without model assumption. In this section, we will discuss how to find out the universal parameterization of moments.

Generally, we are interested in GPDs of (anti)quarks with different flavors as well as the gluons, and conventionally the quarks and antiquark GPDs are collected together by defining the $F_{(q)}(x,\xi,t)$ such that \cite{Muller:2014wxa}
\begin{align}
    F_{q}(x>-\xi,\xi,t)&=F_{(q)}(x,\xi,t)\ ,\\
    F_{\bar q}(x>-\xi,\xi,t)&=-F_{(q)}(-x,\xi,t)\ ,
\end{align}
where $F_{q}(x,\xi,t)$ is the quark GPD and $F_{\bar{q}}(x,\xi,t)$ the antiquark GPD with support $1>x>-\xi$. It is common to reexpress them in terms of the valence and sea distributions analogous to the forward PDF as,
\begin{align}
    F_{q,\rm{val}}(x,\xi,t)&\equiv F_{q}(x,\xi,t)-F_{\bar{q}}(x,\xi,t)\ ,\\
    F_{q,\rm{sea}}&\equiv 2 F_{\bar{q}}(x,\xi,t)\ ,
\end{align}
such that for each quark flavor, the different distributions can be expressed in terms of the corresponding valence and sea distributions.

\subsection{Polynomiality condition and  small $\xi$ expansion of GPD moments}

To impose the physical constraints on GPDs, we start by noting that the polynomiality condition, originally proved for the Mellin moments of the GPDs, can be equivalently applied to the conformal moments. Since each Gegenbauer polynomial $C_{j}^{(\lambda)}(x)$ can be expressed in terms of linear combination of $x^k$,
\begin{equation}
C_{j}^{(\lambda)}(x) =\sum_{k=0}^{j} c^{(\lambda)}_{j,k}x^k\ ,
\end{equation}
and vice versa:
\begin{equation}
\label{eq:meltoconf}
x^{j}= \sum_{k=0}^{j} c^{-1,(\lambda)}_{j,k} C_{k}^{(\lambda)}(x)\ ,
\end{equation}
where $c^{(\lambda)}_{j,k}$ and $c^{-1,(\lambda)}_{j,k}$ are some constant coefficients, according to eq. (\ref{eq:conformalmoment}) we have 
\begin{align}
\begin{split}
    \mathcal F_{j}(\xi,t)&\propto \int_{-1}^{1}\text{d}x ~\xi^{j}  C_{j}^{\frac{3}{2}}\left(\frac{x}{\xi}\right) F(x,\xi,t) \ ,\\
    &=\int_{-1}^{1}\text{d}x  \sum_{k=0}^{j} c^{(\lambda)}_{j,k} \xi^{j-k}x^k F(x,\xi,t) \ ,\\
    &=\sum_{k=0}^{j} c^{(\lambda)}_{j,k} \xi^{j-k} \int_{-1}^{1}\text{d}x   x^k F(x,\xi,t)\ .
\end{split}
\end{align}
Thus the polynomiality condition of Mellin moments immediately leads to the polynomiality condition of conformal moments. The reverse can be proved in the same way using eq (\ref{eq:meltoconf}), so the two polynomiality conditions are equivalent. 
Therefore, we can write the conformal moments of GPDs as 
\begin{equation}
\mathcal F_j(\xi,t)=\sum_{k=0,\rm{ even}}^{j+1} \xi^{k} \mathcal{F}_{j,k}(t)\ ,
\end{equation}
where the form factors $\mathcal{F}_{j,k}(t)$ with double subscripts do not depend on $\xi$. While the GPDs $F(x,\xi,t)$ are completely determined by $\mathcal{F}_{j,k}(t)$, limited information on them is known. Two limit behaviors, the PDF limit $\xi = 0$ and the DA limit $\xi = 1$, can be considered separately. In the semi-forward limit $\xi = 0$, the conformal moments $\mathcal F_j(\xi,t)$ are given by $\mathcal{F}_{j,k}(t)$ with $k=0$:
\begin{equation}
\mathcal F_j(\xi=0,t)= \mathcal{F}_{j,0}(t)\ ,
\end{equation}
where we know that the GPD will be reduced to the PDF and the conformal moments correspond to the Mellin moments of the PDF if we further take $t=0$. Thus, the $\mathcal{F}_{j,0}(t)$ at zero momentum transfer $t=0$ is entirely determined by the forward PDF $f(x)$, which can always be determined well from global fitting, and we have the following requirements
\begin{equation}
\mathcal{F}_{j,0}(t=0)=\int \text{d}x~x^{j} f(x) \ .
\end{equation}
Commonly in the PDF analysis, for instance in the CTEQ global analysis \cite{Hou:2019efy}, the ansatz of the PDF $f(x)$ 
\begin{equation}
f(x)=\sum_{i=1}^{i_{\rm{max}}}N_i x^{-\alpha_i} (1-x)^{\beta_i}\ ,
\end{equation}
is chosen, whereas further improvements can be done based on this. This ansatz can be immediately transformed to the moment space, and we have the following ansatz for the Mellin moments of the PDF 
\begin{equation}
\label{eq:confmoments}
\mathcal{F}_{j,0}(t=0)=\int \text{d}x~x^{j} f(x) = \sum_{i=1}^{i_{\rm{max}}}N_i B(j+1-\alpha_i,1+\beta_i)\ ,
\end{equation}
which will be our building blocks for the GPD at $\xi=0$ or small $\xi$. 

On the other hand, in the DA limit $\xi = 1$, we have 
\begin{equation}
\mathcal F_j(\xi=1,t)=\sum_{k=0,\rm{ even}}^{j+1}  \mathcal{F}_{j,k}(t)\ ,
\end{equation}
which could have completely different behavior than the forward case. Actually, they must behave differently. The reason is that the formal summation in eq. (\ref{eq:conformalsum}) with the moments in eq. (\ref{eq:confmoments}) does not converge, and thus it can only be resummed by the Mellin-Barnes integral method or other equivalent methods. Consequently, GPDs corresponding to such moments do not generally vanish at $|x|=\xi$. While it is expected as GPDs do not generally vanish at the crossover line, for large $\xi$, especially for $\xi$ close to 1, such moments lead to GPDs that do not vanish at the end point $|x|=1$. Therefore, the moments at large $\xi$ must converge faster than the moments in eq. (\ref{eq:confmoments}) in order to have proper end-point behavior. 

Due to the lack of constraints from physically observable to the large $\xi$ behavior of GPDs, we will restrain ourselves to the region with small $\xi$ to avoid such complexity, whereas the large $\xi$ behavior (which also corresponds to the large $t$ behavior as shown in figure \ref{fig:txiplot}) of GPDs requires separate studies. Then, though the moments of the GPDs are generally unknown, in the small $\xi$ region, the first few terms will dominate:
\begin{equation}
\mathcal F_j(\xi,t)=\sum_{k=0,\rm{ even}}^{k_{\rm{cut}}} \xi^{k} \mathcal{F}_{j,k}(t)\ ,
\end{equation}
with $k_{\rm{cut}}$ the cut-off parameter which controls the $\xi$ dependence, and generally larger $k_{\rm{cut}}$ is needed for more flexible $\xi$ dependence to contain the larger $\xi$ region. In the region with small $\xi$, it is assumed that the ansatz used for the PDF in eq. (\ref{eq:confmoments}) applies not only to the leading moment with $k=0$, but also the first few moments with $k=2,4...$ which can certainly break down as $k$ gets large.

Note that the small $\xi$ expansion does not apply to the GPDs themselves, since GPDs are not analytic at the crossover line $|x|=\xi$. However, the moments of GPDs are always polynomials of $\xi$, and they are always regular and expandable in the small $\xi$ region, which, on the other hand, need proper resummation in order to recover the GPDs.

\subsection{Regge trajectory and $t$-dependence of GPDs}

The last piece to complete the universal moment parameterization is the $t$-dependence of GPDs. In the framework of moment expansion, the $t$-dependence is encoded in all the form factors $\mathcal{F}_{j,k}(t)$ and thus the $t$-dependence of the form factors $\mathcal{F}_{j,k}(t)$ are crucial. While each term can be in principle fitted to the experimental measurements or lattice calculations, it is not practical to do so, because the resummation procedure requires a set of $\mathcal{F}_{j,k}(t)$ with $j$ goes to infinity. Neither experimental measurements nor lattice calculations allow us to access all the moments, especially the higher order ones, and thus some extrapolation methods that take the known moments into account and allow us to extrapolate the higher moments will be necessary. Motivated by that, we introduce the Regge trajectory in this subsection and discuss how it can help build the $t$-dependence of GPDs.

\begin{figure}[t]
\centering
\includegraphics[width=0.85\textwidth]{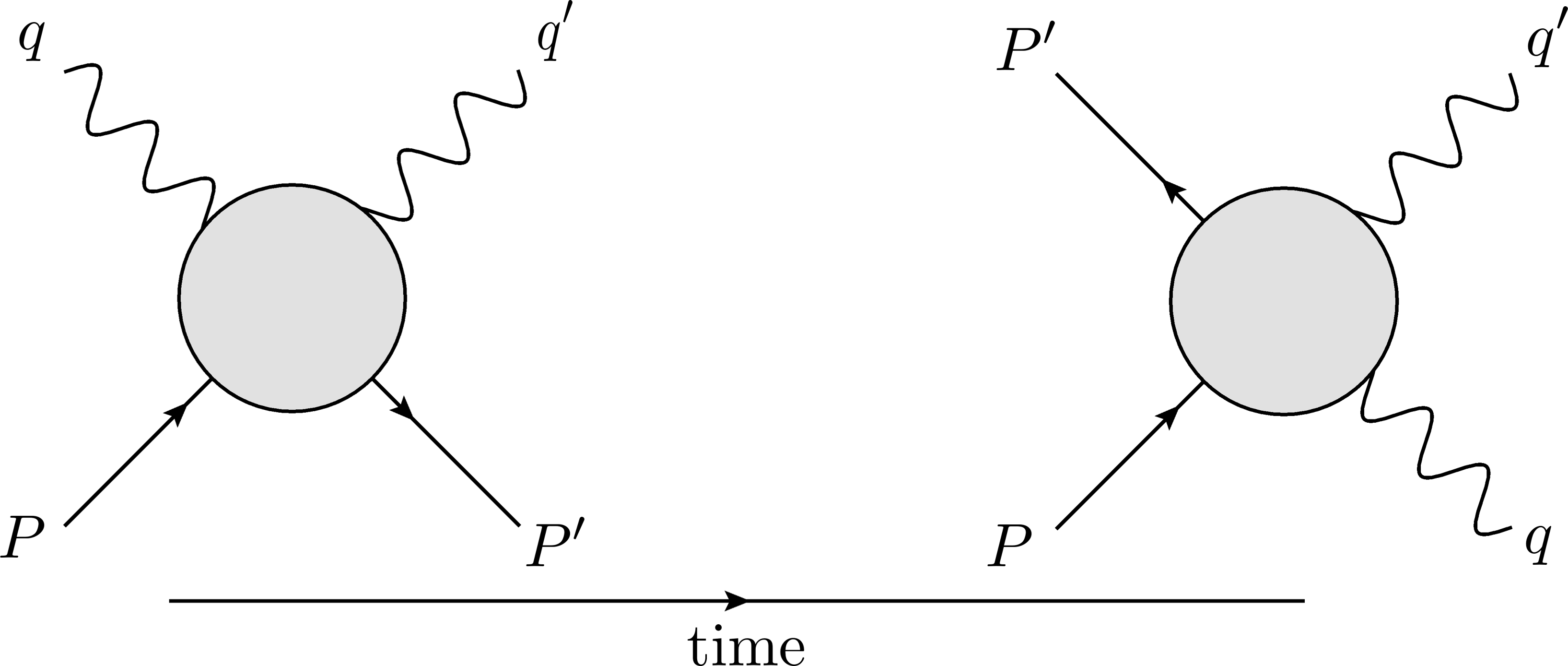}
\caption{\label{fig:crosssymmetry} Crossing symmetry shown in terms of Feynman diagrams. The two processes are related to the same set of Feynman diagrams by simply exchanging the initial and final particles. The right plot corresponds to the $t$-channel process of the left one.}
\end{figure}

The starting point is the crossing symmetry, with which processes of different channels but with the same Feynman diagrams can be connected, as shown in figure. \ref{fig:crosssymmetry}. The scattering amplitude $\mathcal T$ can be written with these Feynman diagrams regardless of which channel one chooses as a function of the Mandelstam variables $s,t,u$, ignoring the spin degrees of freedom for simplicity. The Mandelstam variables $s,t,u$ for $2\to 2$ processes are defined as $s\equiv(P+q)^2,t\equiv(P'-P)^2$ and $u\equiv(P-q')^2$. The 3 Mandelstam variables $s,t,u$ satisfy the condition $s+t+u=P^2+P'^2+q^2+q'^2$, with which we can eliminate the variable $u$ for simplicity.

One key hypothetical property in Regge theory is the duality \cite{Veneziano:1968yb}, which states that the scattering amplitude is given by the sum of all $t$-channel resonances and also equals to the sum of all $s$-channel resonances. Therefore, the $t$-dependence of the scattering amplitude can be tentatively written as 
\begin{equation}
\label{eq:respole}
    \mathcal T(s,t,u)= \sum_{\rm{res.}} \frac{g_{t}(s,u)}{t-m_{\rm{res}}^2}\ ,
\end{equation}
with $g_{t}(s,u)$ the residuals of the resonance poles and the RHS sums over all the $t$-channel resonances. In principle, the scattering amplitude can then be written directly with all the resonances. However, there are potentially infinite resonances and summing over their contributions are highly non-trivial. Thus, one will need to collect the poles together in order to sum over them, which leads to the Regge theory \cite{Regge:1959mz,Regge:1960zc}.

Regge theory is, in short, a theory about complex angular momentum $J$ (different from the conformal spin $j$). The key observation is that the mass of the resonances increases with increasing angular momentum $J$ on the hadron spectrum. Thus, a linear Regge trajectory usually written as $\alpha(t)\equiv \alpha+\alpha' t$ with $\alpha(t)$ the angular momentum and $t$ the mass squared, also known as the Reggeon. The major assumption in Regge theory is that the resonances and their analytic continuation to complex $J$ are the only singularities in the scattering amplitudes. Then the scattering amplitude can be expressed in terms of all the Regge trajectories or Reggeons.

Note that the scattering amplitude $\mathcal T(s,t,u)$ is not explicitly a function of $J$. To separate the contributions of different $J$, it can be written in terms of the partial wave expansion of a $t$-channel scattering as
\begin{equation}
\label{eq:partialwaveexp}
 \mathcal T(s,t,u)= 16\pi \sum_{J=0}^\infty(2J+1)\mathcal T_J(t,u)P_J(\cos\theta_t)\ .
\end{equation}
Then the partial wave amplitudes $\mathcal T_J(t,u)$ will have poles at $J=\alpha(t)$ for each Regge trajectory according to the assumption made for Regge theory, and they can be written with these poles as
\begin{equation}
\label{eq:regpole}
 \mathcal T_J(t,u) = \sum_{\rm{Reg.}} \frac{r(t,u)} {J-\alpha(t)} \ ,
\end{equation}
which sums over all the Regge trajectory and $r(t,u)$ is the corresponding residual. Note that eq. (\ref{eq:respole}) and eq. (\ref{eq:regpole}) are equivalent descriptions of those resonance poles. Their relation is intuitive --- eq. (\ref{eq:respole}) approaches those poles with a fixed integer $J$ in the $M^2$ direction, whereas eq. (\ref{eq:regpole}) approaches those poles with fixed $M^2$ in the $J$ direction.

Then, we can discuss the $t$-dependence of the GPDs or equivalently the form factors $\mathcal F_{j,k}(t)$. Since they are essentially the nucleon matrix elements of certain quark (gluon) operators which are closely related to the quark (gluon)-nucleon scattering amplitude, the above arguments also apply to them, at least qualitatively. Consider the partial wave expansion in eq. (\ref{eq:partialwaveexp}). In the higher energy limit $s\to\infty$ and $\xi \to 0$, the $ \cos(\theta_t)$ can be expressed in terms of $\xi$ as
\begin{equation}
 \cos(\theta_t)\propto -\frac{1}{\xi} \to \infty\ .
\end{equation}
Then with the asymptotic behavior of the Legendre polynomials, 
\begin{equation}
\lim_{z\to\infty} P_J (z) \propto z^{J}\ ,
\end{equation}
one finds that the partial wave amplitude with angular momentum $J$ has the asymptotic behavior of $\xi^{-J}$ in the high energy limit. Then it can be quickly identified that the term $\mathcal F_{j,k}(t)$ corresponds to the angular momentum $J=j+1-k$, which leads to 
\begin{equation}
    \mathcal F_{j,k}(t)= \sum_{\rm{Reg.}} \frac{r_{j,k}(t)}{j+1-k-\alpha(t)}\ ,
\end{equation}
with $r(t)$ the corresponding residuals of the Regge poles. Again, if we stay in the region where momentum transfer $|t|$ is small, the leading lightest Regge trajectory is sufficiently good to describe the $t$-dependence of the GPDs, while more terms can always be added to make it more flexible. Therefore, we reach the finial proposed ansatz for the moments of GPDs as
\begin{equation}
\label{eq:gumpform}
    \mathcal F_{j,k}(t)= \sum_{i=1}^{i_{\rm{max}}}N_{i,k} B(j+1-\alpha_{i,k},1+\beta_{i,k})\frac{r'_{i,j,k}(t)}{j+1-k-\alpha_{i,k}(t)}\ , 
\end{equation}
with $k=0,2,4$ which breaks down at large $k$. Note that the residuals $r'_{i,j,k}(t)$ are smooth functions of $t$. In order to restore the corresponding forward limit in eq. (\ref{eq:confmoments}), it should satisfy $r'_{i,j,k}(t=0)=j+1-k-\alpha_{i,k}$.
However, given the present amount of data, it should be safe to set it to be simply constant:
\begin{equation}
r'_{i,j,k}(t)=j+1-k-\alpha_{i,k}\ ,
\end{equation}
which can be further improved once more data are available. Then there are essentially $4$ parameters for each term in the summation on the RHS of eq (\ref{eq:gumpform}).

In the end, note that the proposed form in eq. (\ref{eq:gumpform}) is equivalent to the ansatz in KM model \cite{Kumericki:2009uq} and the similar proposed forward-like function in the dual parameterization \cite{Muller:2014wxa} except for the $SO(3)$ partial wave expansion that we dropped for simplicity. However, the main goal of this work is to focus on both sea and valence distributions and the corresponding constraints from lattice calculation in for instance \cite{LHPC:2007blg,Hagler:2009ni,Shanahan:2018nnv,Alexandrou:2020zbe,Constantinou:2020hdm,Lin:2020rxa,Lin:2021brq} besides the experimental measurements which only partially constrain the GPDs.

\section{A demonstrative example: $t$-dependent PDF and proton tomography}

\label{sec:tpdf}

Following the above global analysis strategy for GPDs, in this section we show an example to apply it to the $\xi=0$, $t$-dependent PDF and proton tomography from various constraints. Note that once the parameterization is specified, in principle one just needs to put all the constraints together, and try fitting to them with for instance, $\chi^2$ minimization. However, the amount of parameters needed for a global fitting is huge --- we needed at least 4 parameters for each different GPDs ($H$, $E$, $\tilde{H}$ and $\tilde{E}$) and each flavors ($u$, $d$, $g$ and even more) at $\xi=0$ and even more for non-zero skewness $\xi$. Therefore, as a proof of principle we start by considering the $H$ and $E$ GPDs of $u$ and $d$ quarks at $\xi=0$, which are the so-called $t$-dependent PDF $H_{u/d}(x,t)$ and $E_{u/d}(x,t)$. These two quantities are necessary to understand the angular momentum densities of $u$ and $d$ quarks \cite{Ji:1996ek}
\begin{equation}
\label{eq:quarkamdensity}
    J_{u/d}(x,t)=\frac{1}{2}x\left(H_{u/d}(x,t)+E_{u/d}(x,t)\right)\ ,
\end{equation}
which will be our main focus in this section.

\subsection{Fitting the valence $t$-dependent PDF}

\begin{figure}[t]
\centering
\begin{minipage}[b]{\textwidth}
\centering
\includegraphics[width=0.8\textwidth]{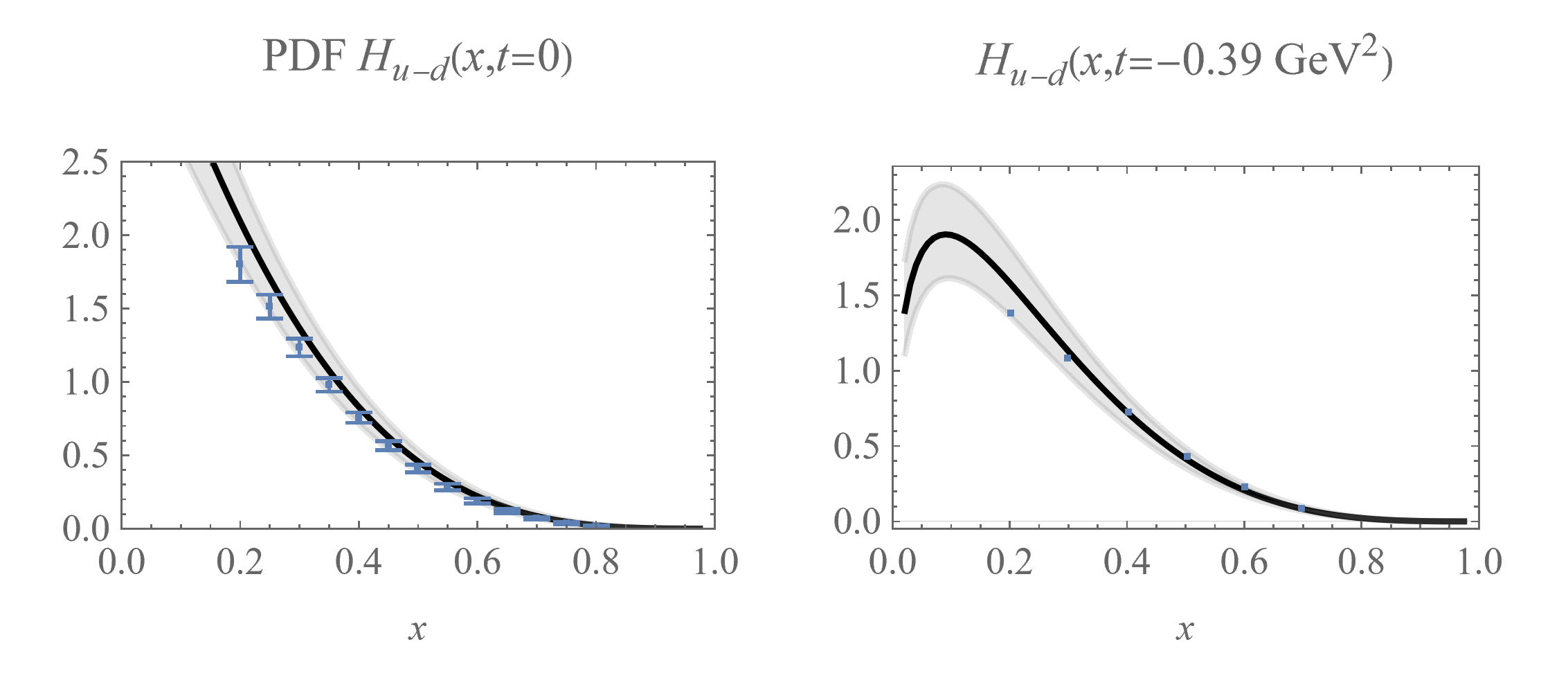}
\end{minipage}
\begin{minipage}[b]{\textwidth}
\centering
\includegraphics[width=\textwidth]{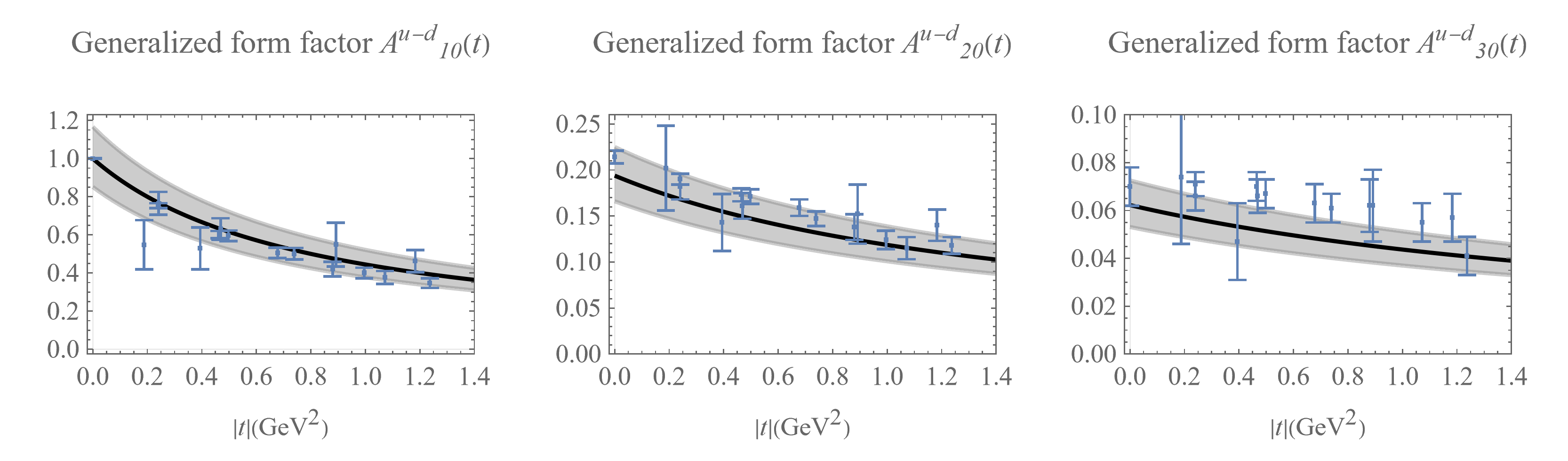}
\end{minipage}
\caption{\label{fig:hvecfit} A simultaneous fit to the forward PDF $H_{u-d}(x)$ (corresponding to the PDF $q_{u-d}(x)$) \cite{Hou:2019efy,Xie:2021equ}, lattice calculated PDF $H_{u-d}(x,t=-0.39 \text{ GeV}^2)$ \cite{Lin:2020rxa} and generalized form factors $A^{u-d}_{10}(t)$, $A^{u-d}_{20}(t)$ and $A^{u-d}_{30}(t)$ \cite{LHPC:2007blg}. The solid lines are the central value of $H_{u-d}(x,t)$, the gray bands are the regions within 1 $\sigma$ uncertainty and the dots are measurements that are fitted to. The $\chi^2$ per d.o.f. is 3.0 for our 4 parameters fitting. The parameters from the fitting read $N_{H\rm{vec}}=4.518~(327)$, $\alpha_{H\rm{vec}}=-0.027~(26)$, $\beta_{H\rm{vec}}=3.269~(76)$ and $\alpha'_{H\rm{vec}}=1.286~(51)$.}
\end{figure}

In this subsection, we use the proposed form in eq. (\ref{eq:gumpform}) for the $t$-dependent PDF $H_{u/d}(x,t)$ and $E_{u/d}(x,t)$. We take $k=0$ in the semi-forward $\xi=0$ limit, and set $i_{\rm{max}}=1$ for the simplest ansatz. Since in lattice QCD the flavor singlet $u+d$ and non-singlet $u-d$ combinations are commonly separately calculated, we apply such ansatz for the $H/E_{u+d}(x,t)$ and $H/E_{u-d}(x,t)$ respectively.

We will neglect the sea contributions in the fitting to lattice results here since the sea contributions to GFFs are suppressed compared to the valence contributions. According to the CTEQ18 PDFs \cite{Hou:2019efy}, the moments of sea distributions are about one order of magnitude smaller than the valence ones, where similar suppression is expected for the GFFs. Therefore, one should not expect strong constraints to the sea distributions from lattice results, and the experimental measurements are crucial to get the sea distributions.

\begin{figure}[t]
\centering
\begin{minipage}[t]{\textwidth}
\centering
\includegraphics[width=0.9\textwidth]{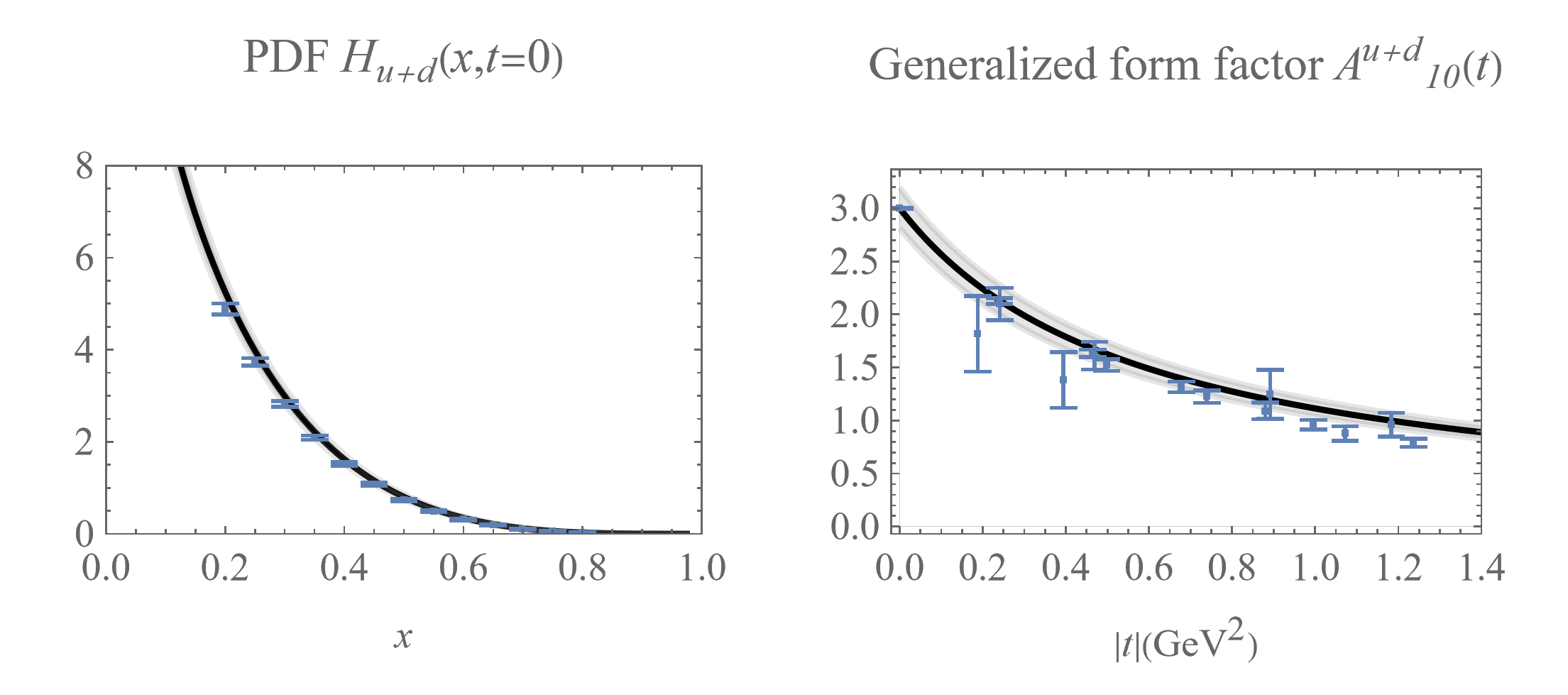}
\end{minipage}
\begin{minipage}[t]{\textwidth}
\centering
\includegraphics[width=0.9\textwidth]{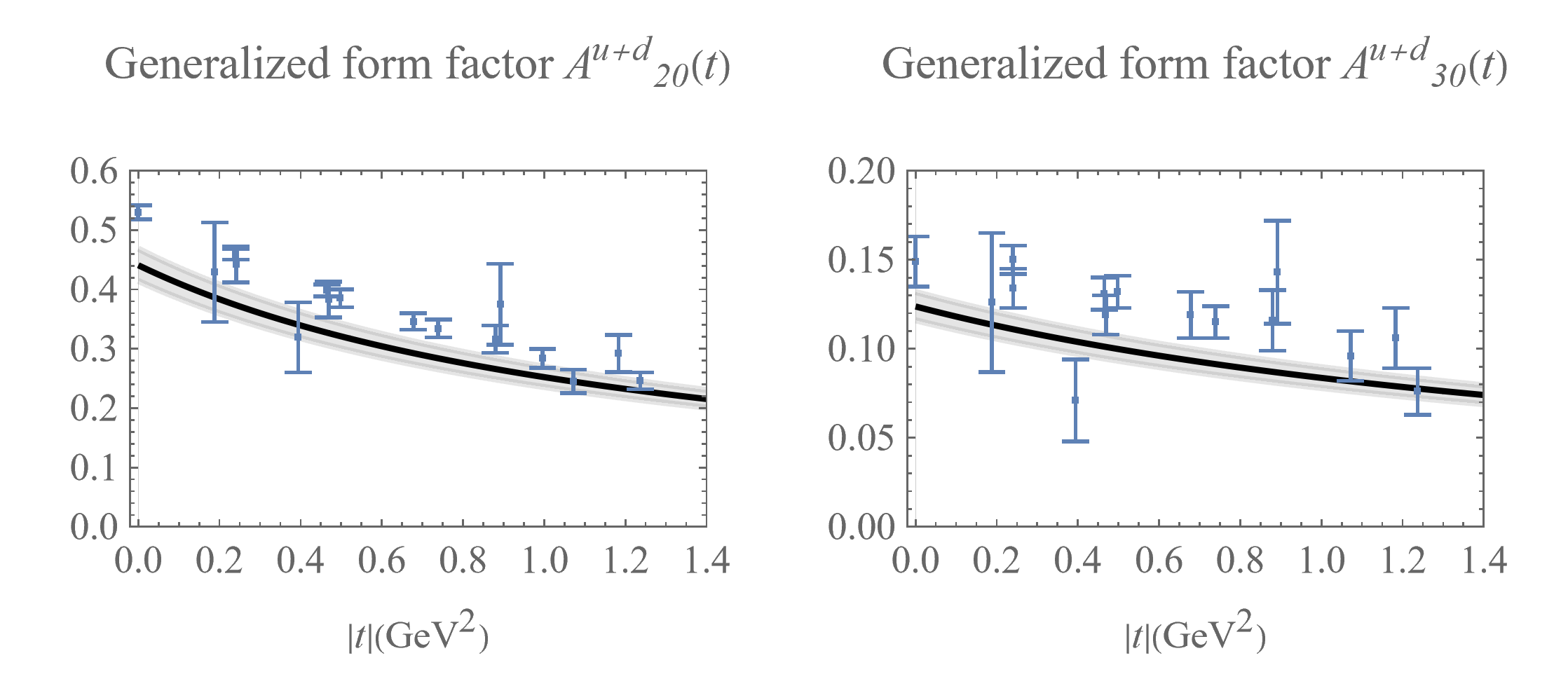}
\end{minipage}
\caption{\label{fig:hscalfit} A simultaneous fit to the forward PDF $H_{u+d}(x)$ (corresponding to the PDF $q_{u+d}(x)$) \cite{Hou:2019efy,Xie:2021equ} and generalized form factors $A^{u+d}_{10}(t)$, $A^{u+d}_{20}(t)$ and $A^{u+d}_{30}(t)$ \cite{LHPC:2007blg}. The solid lines are the central value of $H_{u+d}(x,t)$, the gray bands are the regions within 1 $\sigma$ uncertainty and the dots are measurements that are fitted to. The $\chi^2$ per d.o.f. is 8.47 for our 4 parameters fitting. The parameters from the fitting read $N_{H\rm{scal}}=8.340~(263)$, $\alpha_{H\rm{scal}}=0.210~(10)$, $\beta_{H\rm{scal}}=3.584~(43)$ and $\alpha'_{H\rm{scal}}=1.338~(27)$.}
\end{figure}

\begin{figure}[t]
\centering
\begin{minipage}[b]{\textwidth}
\centering
\includegraphics[width=0.9\textwidth]{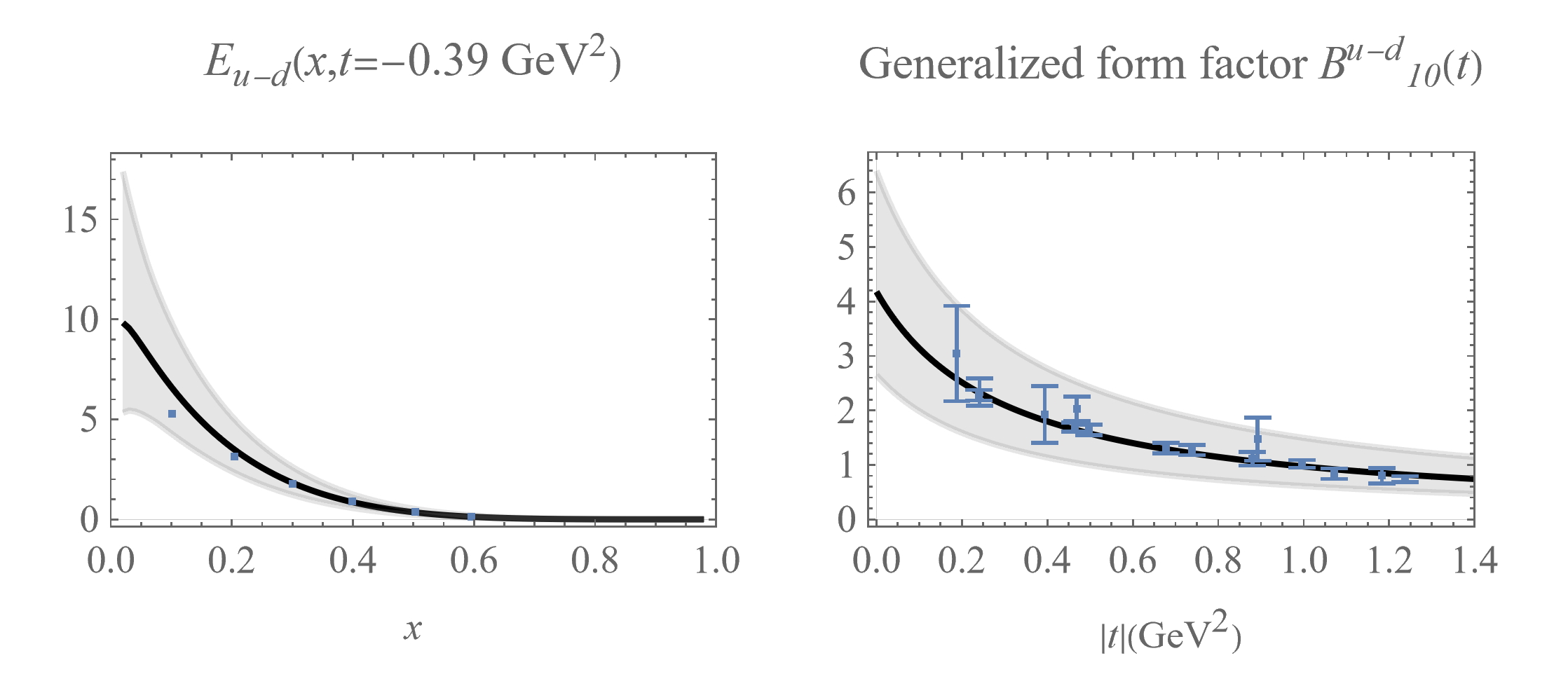}
\end{minipage}
\begin{minipage}[b]{\textwidth}
\centering
\includegraphics[width=0.9\textwidth]{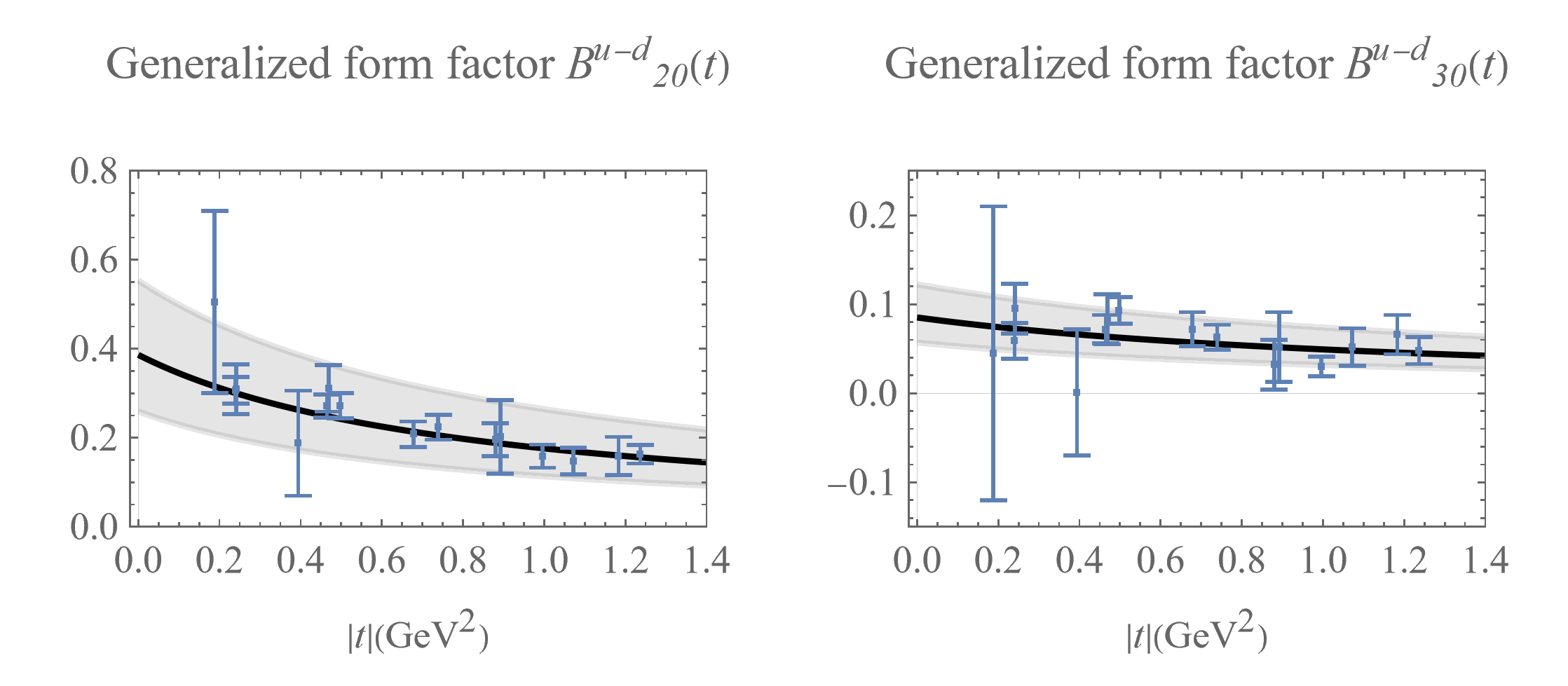}
\end{minipage}
\caption{\label{fig:evecfit} A simultaneous fit to the $t$-dependent PDF $E_{u-d}(x)$ \cite{Lin:2020rxa} and generalized form factors $B^{u-d}_{10}(t)$, $B^{u-d}_{20}(t)$ and $B^{u-d}_{30}(t)$ \cite{LHPC:2007blg}. The solid lines are the central value of $E_{u-d}(x,t)$, the gray bands are the regions within 1 $\sigma$ uncertainty and the dots are measurements that are fitted to. The $\chi^2$ per d.o.f. is 0.8 for our 4 parameters fitting. The parameters from the fitting read $N_{E\rm{vec}}=6.770~(1.354)$, $\alpha_{E\rm{vec}}=0.437~(62)$, $\beta_{E\rm{vec}}=4.527~(240)$ and $\alpha'_{E\rm{vec}}=1.864~(222)$.}
\end{figure}

For the $t$-dependent PDF $H^{\rm{val}}_{u/d}(x,t)$, we can simultaneously fit to the forward PDF $q(x)$, for which we choose the recommended CTEQ18qed results \cite{Hou:2019efy,Xie:2021equ}, the quark GFFs \cite{LHPC:2007blg} and lattice calculated isovector $t$-dependent PDF $H^{\rm{val}}_{u-d}(x,t)$ \cite{Lin:2020rxa}. We used the ManeParse \cite{Clark:2016jgm} Mathematica package to read the CTEQ18qed PDF and the MINUIT2 minimizer \cite{Hatlo:2005cj} for $\chi^2$ analysis, and the results are given in figure \ref{fig:hvecfit}, where the generalized form factors are defined by \cite{Ji:1996ek,Ji:1998pc}
\begin{equation}
    \begin{aligned}
\int_{-1}^{+1} d x x^{n-1} H\left(x, \xi, t\right)&= \sum^{n-1}_{i=0,\rm{even}}(-2 \xi)^{i} A_{n i}\left(t\right)+\left.(-2 \xi)^{n} C_{n 0}\left(Q^{2}\right)\right|_{n \text { even }}\ ,\\
\int_{-1}^{+1} d x x^{n-1} E\left(x, \xi, t\right)&= \sum^{n-1}_{i=0,\rm{even}}(-2 \xi)^{i} B_{n i}\left(t\right)-\left.(-2 \xi)^{n} C_{n 0}\left(Q^{2}\right)\right|_{n \text { even }}\ ,
\end{aligned}
\end{equation}
thus only $A_{n0}(t)$ and $B_{n0}(t)$ survive in the semi-forward limit $\xi=0$. Note that the $\chi^2$ per degree of freedom (d.o.f.) is 3.0 for our 4 parameters fitting. The major source of $\chi^2$ is the discrepancy between lattice calculated form factors and the global fitted PDFs --- the moments of PDF from global fitting are significantly lower than the lattice calculations. Therefore, the $\chi^2$ per d.o.f. indicates the systematic uncertainties including the effects of unphysical pion mass of $m_{\pi}= 496  \text{ MeV}$, omitted disconnected diagrams and other effects in lattice calculation that are not taken into account.

Similar fitting can be done for the isoscalar valence distributions $H^{\rm{val}}_{u+d}(x,t)$ as shown in figure \ref{fig:hscalfit}. However, since the isoscalar $t$-dependent PDF $H^{\rm{val}}_{u+d}(x,t)$ is not calculated in ref. \cite{Lin:2020rxa}, we fit to the forward PDF $H^{\rm{val}}_{u+d}(x,t=0)$ and the lattice isoscalar form factors $A^{u+d}_{10}(t)$, $A^{u+d}_{20}(t)$ and $A^{u+d}_{30}(t)$ \cite{LHPC:2007blg}.  Again, the major source of $\chi^2$ is the discrepancy between lattice calculated form factors and the global fitted PDFs which do not indicate the failure of the parameterization form used and can be mitigated if the systematical uncertainties are under control.

For the $t$-dependent PDF $E^{\rm{val}}_{u/d}(x,t)$, the forward constraints from PDFs no longer exist, and we have to rely on the lattice calculated $t$-dependent GPDs $E^{\rm{val}}_{u/d}(x,t)$ \cite{Lin:2020rxa}. Though the isoscalar GPD $E_{u+d}(x,t)$ is not calculated there, there is strong evidence showing that the isoscalar form factors $B_{u+d}(t)$s are consistent with zero \cite{LHPC:2007blg,Shanahan:2018nnv}. Therefore, we will assume $E^{\rm{val}}_{u+d}(x,t)=0$ with which the flavor separation becomes trivial. The results for the $E^{\rm{val}}_{u-d}(x,t)$ are shown in figure \ref{fig:evecfit}. The $\chi^2$ per d.o.f. is about 0.8, which is really due to the lack of precision in the lattice results, since there are no constraints from better determined forward PDFs. Therefore, the uncertainties of the GPD $E^{\rm{val}}_{u-d}(x,t)$ are significantly larger.

\subsection{The angular momentum sum rule and proton tomography}

\begin{figure}[t]
\centering
\includegraphics[width=0.6\textwidth]{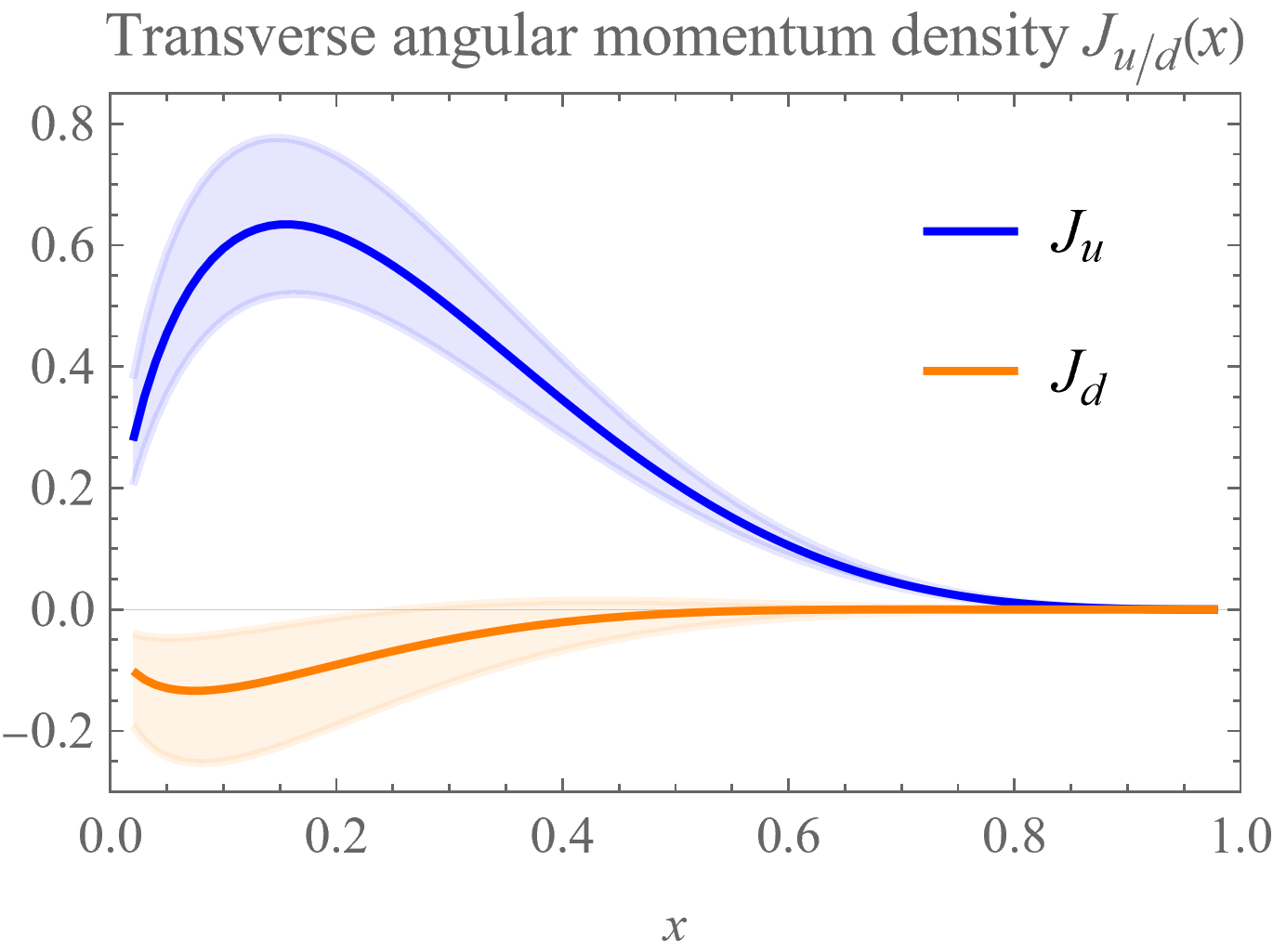}
\caption{\label{fig:judplot} A plot of the quark transverse angular momentum densities $J_{u/d}(x)$ with the fitted GPDs $H_{u/d}(x,t)$ and $E_{u/d}(x,t)$. Most of the uncertainties are from the GPDs $E_{u/d}(x,t)$ which are less constrained without corresponding forward PDFs.}
\end{figure}

With the fitted $t$-dependent PDFs $H_{u/d}(x,t)$ and $E_{u/d}(x,t)$, interesting information of the nucleon 3D structure can be revealed.
On the one hand, as mentioned at the beginning of the section, these GPDs carry information about the transverse angular momentum densities of quarks. On the other hand, as first pointed out by M. Burkardt in refs. \cite{Burkardt:2000za,Burkardt:2002hr}, the GPDs in the $\xi\to 0$ limit can be Fourier transformed in the transverse space to get the impact parameter space parton distribution which describe the 3D structure of the nucleon. In this section, we will explore the proton 3D structure with the above fitted GPDs.

In figure \ref{fig:judplot}, we plot the transverse angular momentum densities of quark with eq. (\ref{eq:quarkamdensity}). Note again that, this plot needs to be further improved with the sea distributions, whose contributions to the angular momentum are, however, mostly in the small $x$ region and one order of magnitude smaller than the valence ones. The contributions from the $d$ quark are always negative according to the fitted results, which are essentially the consequence of the assumption $E_{u+d}(0)=0$ supported by the lattice calculations of quark and gluon generalized form factors. 

\begin{figure}[t]
\centering
\includegraphics[width=1\textwidth]{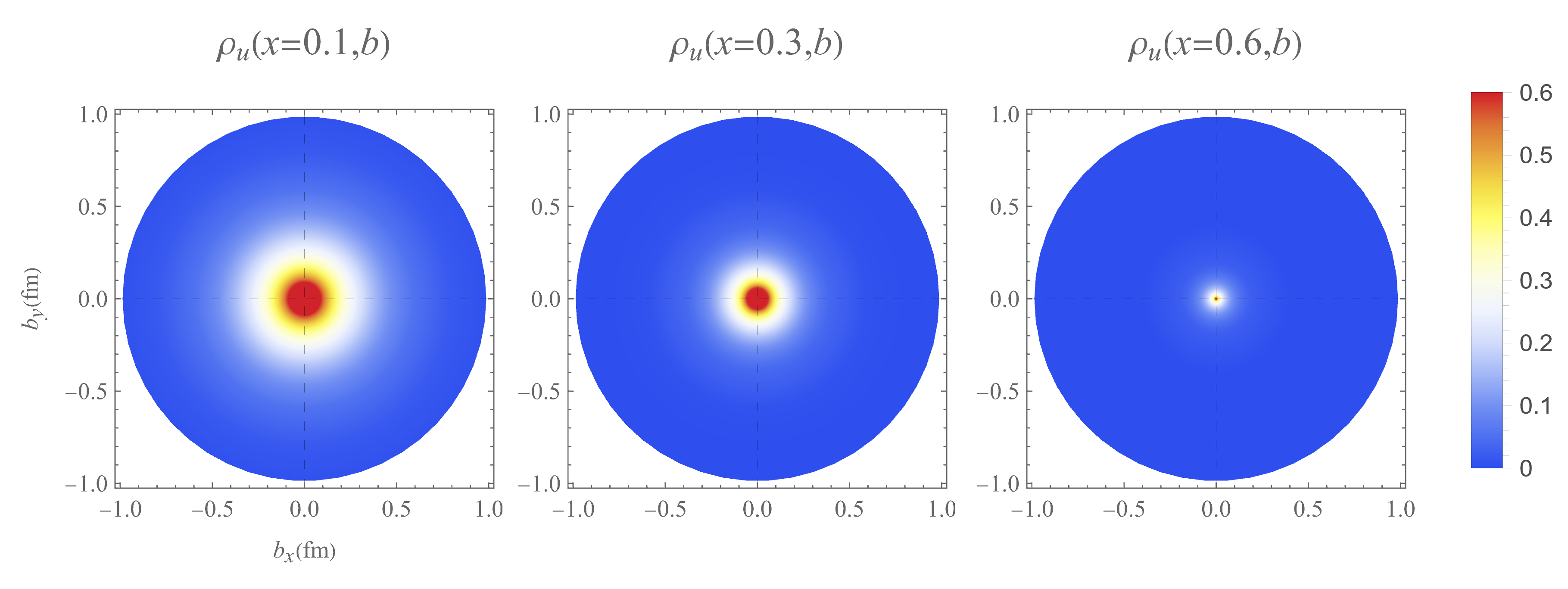}
\begin{minipage}[b]{\textwidth}
\centering
\includegraphics[width=1\textwidth]{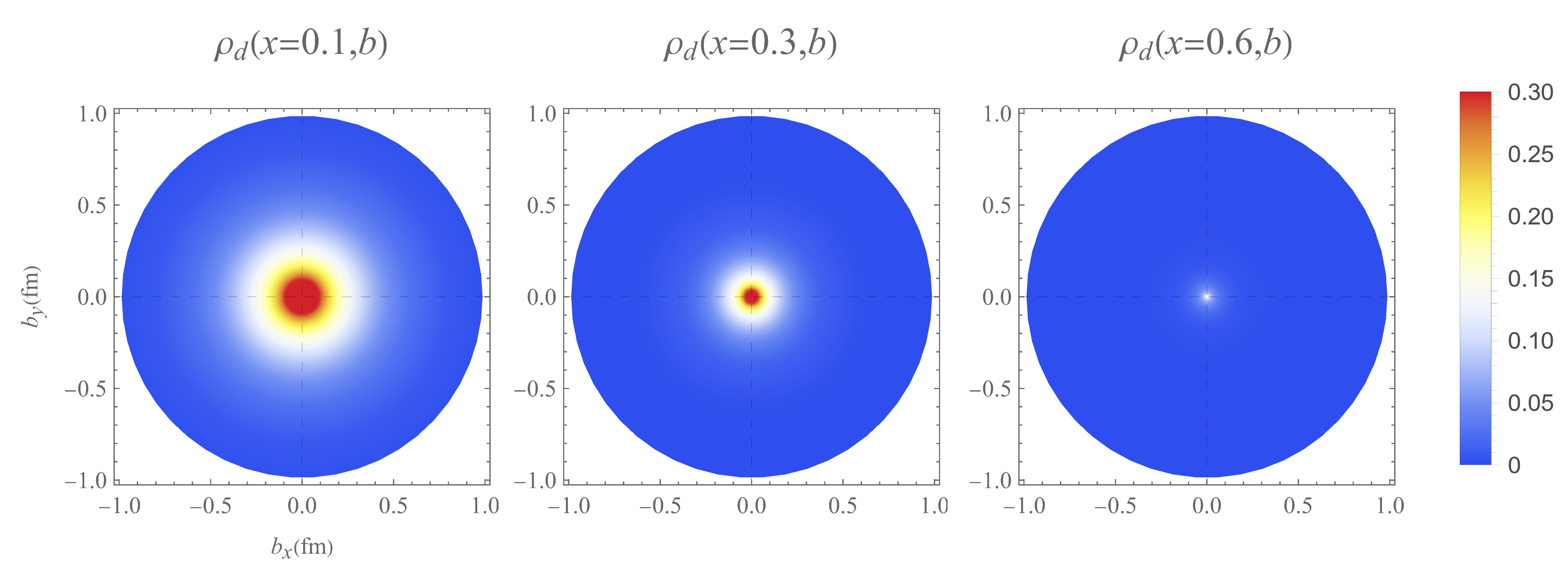}
\end{minipage}
\caption{\label{fig:qdensitytomography} Plots of quark densities $\rho_{q}(x,\boldsymbol b)$ for both $u$ (the upper row) and $d$ (the lower row) quarks in unpolarized proton with increasing momentum fraction $x$. The densities typically get more concentrated but suppressed for increasing $x$. }
\end{figure}

In addition to the parton transverse angular momentum densities which only make use of the GPD in the forward limit $\xi=t=0$, we can also take advantage of the $t$-dependent of the GPDs and study the impact parameter space distribution of the quarks. For instance, the quark densities in unpolarized protons are given by the Fourier transform of the GPDs $H_{u/d}(x,t)$:
\begin{equation}
\label{eq:unpdensity}
\begin{split}
    \rho_{q}(x,\boldsymbol b) = \int \frac{\text{d}^2\boldsymbol \Delta}{(2\pi)^2} e^{-i \boldsymbol{\Delta}\cdot \boldsymbol b}H_q(x,-\boldsymbol \Delta^2)    =\mathscr H_{q}(x,\boldsymbol b)\ ,
\end{split}
\end{equation}
where $\mathscr H_{q}(x,\boldsymbol b)$ is defined as the 2-dimensional Fourier transformation of $H_q(x,-\boldsymbol \Delta^2)$. Then we plot the quark distribution with the above expression and the fitted GPDs in figure \ref{fig:qdensitytomography}. The quark densities typically get more concentrated but suppressed for increasing $x$, since the number densities should decrease for quarks with larger momentum due to momentum conservation, and the less deflected they will be.

\begin{figure}[t]
\centering
\includegraphics[width=1\textwidth]{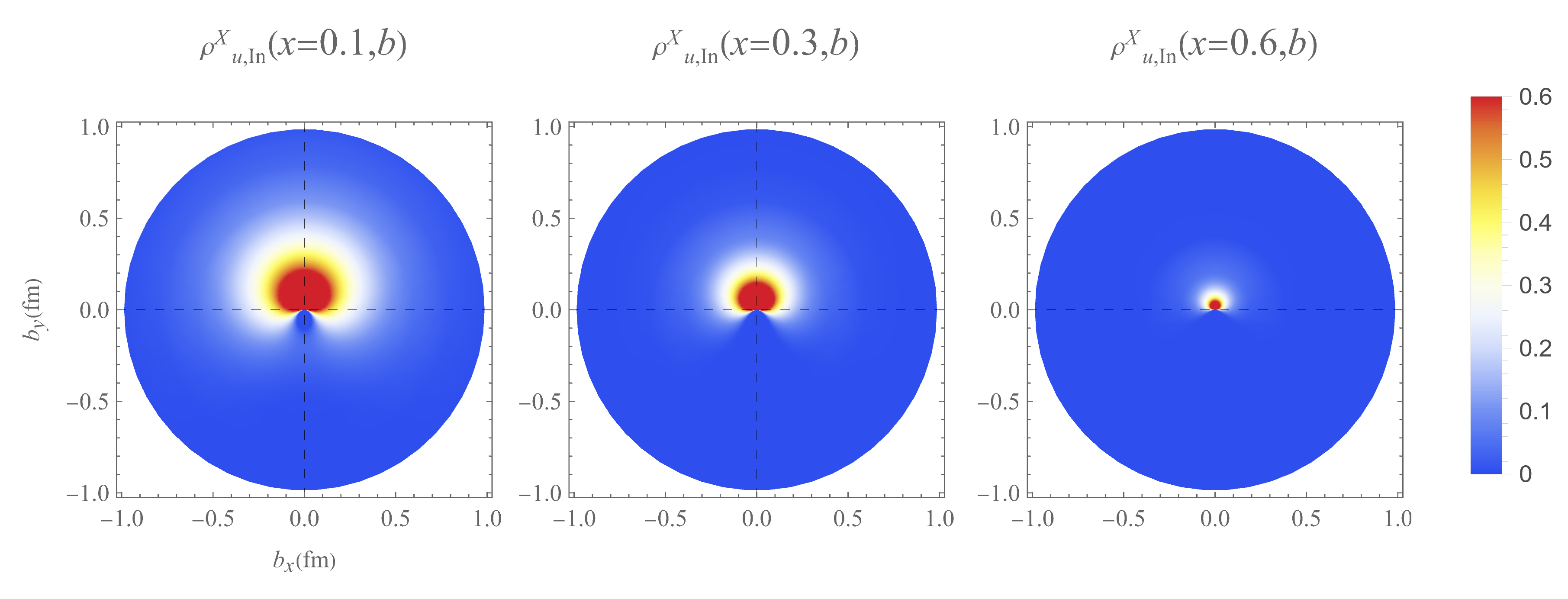}
\begin{minipage}[b]{\textwidth}
\centering
\includegraphics[width=1\textwidth]{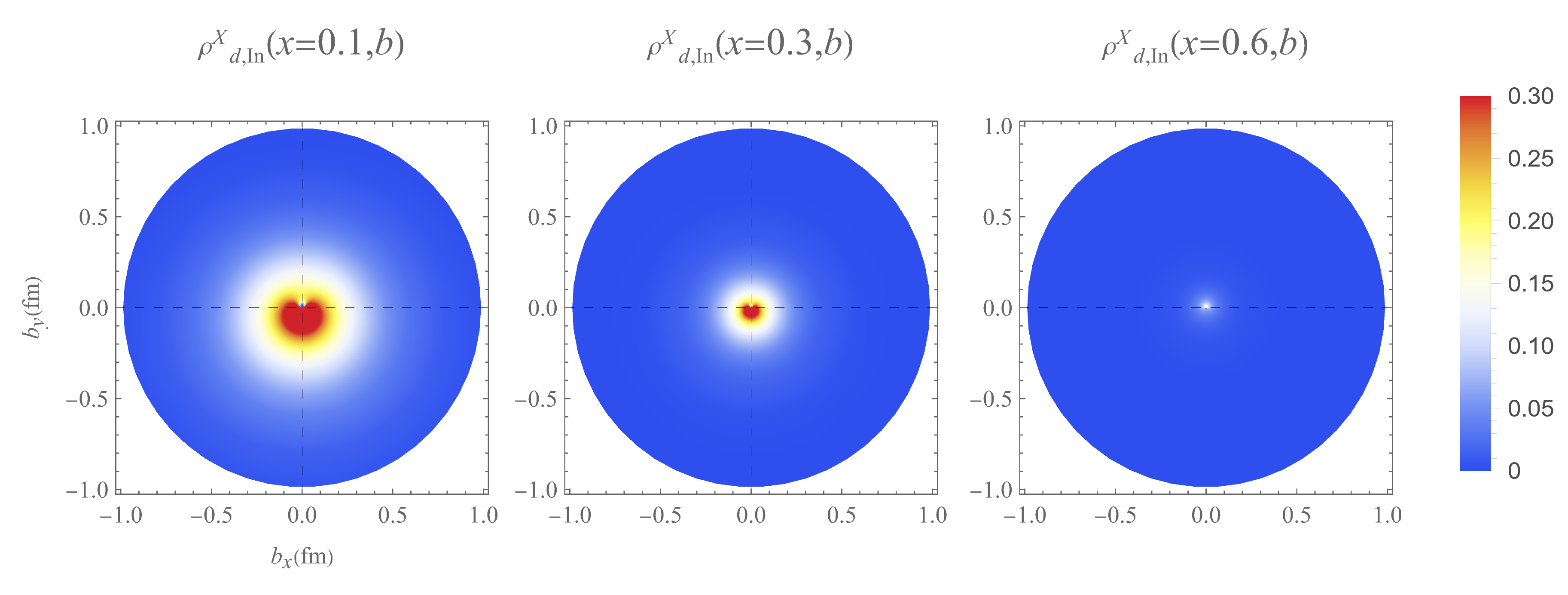}
\end{minipage}
\caption{\label{fig:qTdensitytomography} Plots of the intrinsic quark densities $\rho_{q,\rm{In}}^{X}(x,\boldsymbol b)$ for both $u$ (the upper row) and $d$ (the lower row) quarks in transversely polarized proton (in the $X$ direction) with increasing momentum fraction $x$. Both the $u$ and $d$ quark densities are shifted in the $y$ direction which contribute to the transverse angular momentum $J^{X}$, while the $u$ contributions are positive ($+y$ direction) and the $d$ contributions are negative ($-y$ direction).}
\end{figure}

Another set of densities that is of interest is the quark densities in a polarized proton. It is worth noting that the transverse angular momentum densities defined in eq. (\ref{eq:quarkamdensity}) are interpreted as the parton angular momentum densities in a transversely polarized nucleon only as pointed out in ref. \cite{Burkardt:2002hr}, see also  \cite{Burkardt:2005hp,Ji:2012sj,Ji:2012ba}. The original quark density in a transversely polarized nucleon defined in ref. \cite{Burkardt:2002hr} reads
\begin{equation}
\begin{split}
\label{eq:rhoXb}
    \rho_{q}^{X}(x,\boldsymbol b)=&\int \frac{\text{d}^2\boldsymbol \Delta}{(2\pi)^2} e^{-i \boldsymbol{\Delta}\cdot \boldsymbol b} \left(H_q(x,-\boldsymbol \Delta^2)+\frac{i\Delta_y}{2M}E_q(x,-\boldsymbol \Delta^2)\right)\ ,\\
    =& \mathscr H_q(x,\boldsymbol b) -\frac{1}{2M}\frac{\partial}{\partial b^y} \mathscr E_q(x,\boldsymbol b)\ ,
\end{split}
\end{equation}
with $\mathscr E_q(x,\boldsymbol b)$ the 2-dimensional Fourier transformation of $E_q(x,-\boldsymbol \Delta^2)$ and the nucleon polarized in the $X$ (distinguished from the parton momentum $x$) direction. However, with such quark densities, one find
\begin{equation}
\begin{split}
\label{eq:sumrhoXb}
    J_q^{X}(x) \not = \int \text{d}^2\boldsymbol b ( b^y \times  x P^+)\rho_{q}^{X}(x,\boldsymbol b)\ ,
\end{split}
\end{equation}
so the transverse angular momentum densities can not be recovered this way. It is later found that this is due to the nucleon's center of mass motion --- the center of the nucleon has transverse displacement when transversely polarized and thus the longitudinal motion of the whole nucleon will contribute to the transverse angular momentum which does NOT correspond to the intrinsic angular momentum (spin) of the nucleon. This has been discussed in ref. \cite{Ji:2020hii} and carefully studied in ref. \cite{Guo:2021aik} where the intrinsic quark densities are defined as
\begin{equation}
\label{eq:rhoInXb}
\begin{split}
    \rho_{q,\rm{In}}^{X}(x,\boldsymbol b)=&\int \frac{\text{d}^2\boldsymbol \Delta}{(2\pi)^2} e^{-i \boldsymbol{\Delta}\cdot \boldsymbol b} \left[H_q(x,-\boldsymbol \Delta^2)+\frac{i\Delta_y}{2M}\left(H_q(x,-\boldsymbol \Delta^2)+E_q(x,-\boldsymbol \Delta^2)\right)\right]\ , \\
    =&  \mathscr H_q(x,\boldsymbol b)-\frac{1}{2M}\frac{\partial}{\partial b^y} \left( \mathscr H_q(x,\boldsymbol b)+ \mathscr E_q(x,\boldsymbol b) \right) \ ,\\
    =&  \mathscr H_q(x,\boldsymbol b)-\frac{1}{2M}\frac{b_y}{ |\boldsymbol b|}\frac{\partial}{\partial |\boldsymbol b|} \left( \mathscr H_q(x,\boldsymbol b)+ \mathscr E_q(x,\boldsymbol b) \right) \ ,
\end{split}
\end{equation}
with which one has 
\begin{equation}
\begin{split}
\label{eq:sumrhoInXb}
    J_q^{X}(x)  = \int \text{d}^2\boldsymbol b ( b^y \times  x P^+)\rho_{q,\rm{In}}^{X}(x,\boldsymbol b)\ .
\end{split}
\end{equation}
With the above fitted GPDs $H_{u/d}(x,t)$ and $E_{u/d}(x,t)$, the intrinsic quark densities in transversely polarized nucleon and be studied as well as shown in figure \ref{fig:qTdensitytomography}. While the quark densities in a polarized proton get more concentrated but suppressed for increasing $x$ just like in the unpolarized case, we find that clearly the quarks have non-zero transverse displacement, which contributes to the transverse angular momentum of the nucleon along with their large longitudinal momenta.

\subsection{Sea distributions and experimental observables}

Here we discuss the effects of the sea distributions, which are important sources of systematic uncertainties in the above analysis, since in principle both sea and valence distributions contribute to the generalized form factors. As mentioned above, the sea distributions can be dropped because they are mostly in the small $x$ regions and their contributions to the form factors are suppressed compared to that of the valence distributions. On the other hand, this indicates that even if we consider the sea distributions in the above analysis, they will not be constrained as much. Therefore, it is crucial to also include the experimental observables, especially the ones with small $\xi$ (since it can not be zero) in order to constrain the sea distributions. Then, we are inevitably faced with the effects of both none-zero skewness $\xi$ and the sea distributions simultaneously.

Fortunately, in the KM model \cite{Kumericki:2009uq}, they show that the sea distributions with such parameterization method can successfully fit to the experimental data which proves the flexibility of this parameterization when extending to non-zero $\xi$. In this sense, our results here with the valence distributions are complementary to the sea distributions obtained in the KM models. On the other hand, GPDs at none-zero $\xi$ are faced with the inverse problem \cite{Bertone:2021yyz}, as the experimental measurements of CFFs alone cannot determine the $x$-dependence at non-zero $\xi$. The solution (compromise) in the KM model is to model the $x$-dependence of the GPDs at non-zero $\xi$ with the same parameters as the PDFs, except for a different overall normalization constant, which might be too constraining. However, there have been results of lattice calculated $x$-dependence of GPDs with LaMET \cite{Alexandrou:2020zbe} at none-zero $\xi$. These results, even just qualitatively, can be very helpful in mitigating or resolving the inverse problem at non-zero $\xi$.

Therefore, to obtain the GPDs at non-zero $\xi$ reliably, we need to extend our present results with sea distributions and fit to both experimental observables and lattice results at none-zero $\xi$ as well. A global fitting program is then necessary for this purpose, which will be studied in the future coming paper.

\section{Conclusion}

\label{sec:conc}

We present a flexible parameterization program for GPDs employing the universal moment expansion which can be fitted to the forward PDFs, the lattice calculation and the experimental measurements when the skewness parameter $\xi$ is small. Under such parameterization, the polynomiality conditions of GPDs can be easily imposed and the non-analyticity at $x=\xi$ will be taken care of. Since the leading order QCD evolution is multiplicative in the conformal moment space, the QCD scale evolution can be performed easily. 

We briefly discuss the major parameterization methods of GPDs in the literatures and introduce the necessary techniques for building the GPDs in moment space. Following the previous works on parameterizing GPDs with moments, the $t$-dependence of GPDs is parameterized with the Regge theory which shows reasonable agreement with lattice calculated generalized form factors. Then by expanding the moments as powers of $\xi$, the $\xi$-dependence of GPDs can be gradually extended and fitted to observables. The $x$-dependence of GPDs is the least known except in the forward limit, which needs further inputs, most likely from lattice QCD.

As a proof of principle, we apply our fitting program to construct the quark $t$-dependent PDFs $H_{u/d}(x,t)$ and $E_{u/d}(x,t)$ by fitting to the forward PDFs, lattice calculated $t$-dependent PDFs and lattice calculated generalized form factors. With these fitted results, we study the quark transverse angular momentum densities as well as the impact parameter space distributions of quarks in both unpolarized and transversely polarized protons, and show that the quark densities in transversely polarized nucleons are shifted in the transverse direction of impact parameter space, such that the transverse angular momentum can be generated. We also comment on the future development of this program to include sea distributions, non-zero skewness $\xi$ and experimental observables. 

\section*{Acknowledgments}

We thank B. Kriesten and M. Gabriel Santiago for discussions related to the subject of this paper. This research is supported by the U.S. Department of Energy, Office of Science, Office of Nuclear Physics, under contract number DE-SC0020682, and the Center for Nuclear Femtography, Southeastern Universities Research Association, Washington D.C. Y. Guo is partially supported by the JSA/JLab Graduate Fellowship.

\newpage 

\bibliographystyle{jhep}

\bibliography{refs.bib}

\end{document}